# Pentagon, Hexagon, or Bridge? Identifying the Location of a Single Vanadium Cation on Buckminsterfullerene Surface

Jianzhi Xu[1], Joost M. Bakker[2], Olga V. Lushchikova[2], Peter Lievens[3], Ewald Janssens[3*], Gao-Lei Hou[1*]

1. MOE Key Laboratory for Non-Equilibrium Synthesis and Modulation of Condensed Matter, School of Physics, Xi´an Jiaotong University, Xi´an, 710049 Shaanxi, China

2. Radboud University, HFML-FELIX, Toernooiveld 7, 6525 ED Nijmegen, The Netherlands

3. Quantum Solid-State Physics, KU Leuven, Celestijnenlaan 200D, 3001 Leuven, Belgium

*Email: gaolei.hou@xjtu.edu.cn (G.-L.H.) ; ewald.janssens@kuleuven.be (E.J)

**ABSTRACT:** Buckminsterfullerene $C_{60}$ has received extensive research interest ever since its discovery. In addition to its interesting intrinsic properties of exceptional stability and electron-accepting ability, the broad chemical tunability by decoration or substitution on the $C_{60}$-fullerene surface makes it a fascinating molecule. However, to date there is uncertainty about the binding location of such decorations on the $C_{60}$ surface, even for a single adsorbed metal atom. In this work, we report the gas-phase synthesis of the $C_{60}V^+$ complex and its in-situ characterization by mass spectrometry and infrared spectroscopy with the help of quantum chemical calculations and molecular dynamics simulations. We identify the most probable binding position of a vanadium cation on $C_{60}$ above a pentagon center in $\eta^5$-fashion, demonstrate a high thermal stability for this complex, and explore the bonding nature between $C_{60}$ and the vanadium cation, revealing that large orbital and electrostatic interactions lie at the origin of the stability of the $\eta^5$-$C_{60}V^+$ complex.

## Introduction

$C_{60}$ was discovered by Smalley, Kroto, Curl, and co-workers in the mid-1980s and was recognized as the third allotrope of carbon besides graphite and diamond.[1, 2] Its discovery has opened the age of fullerene science and sparked worldwide interest, demonstrated by the awarding of the Noble prize in Chemistry in 1996. This football-shaped molecule, consisting of 20 hexagons and 12 pentagons with icosahedral ($I_h$) symmetry, has interesting intrinsic properties of high stability and electron-accepting ability. In addition, fullerenes show broad structural tunability by decoration or substitution, either on their exterior surface or interior cavity, making them exceptional materials to explore.[2, 3] In particular, significant interest in metal doped fullerenes has emerged just a few years after the discovery of $C_{60}$. For instance, alkali metal doped fullerene films were found to present superconducting behavior,[4] alkali and alkaline earth metal doped fullerenes have high hydrogen storage capability,[5, 6] transition metal doped fullerenes show interesting applications in organic photovoltaic cells,[7, 8] and hybrid structures of $C_{60}$-fullerene with coinage metals can be used in novel nanoscale devices.[9, 10] Each of these phenomena are to some extent linked to electron transportation in hybrid fullerene-metal junctions.[7, 10-13]

However, there is to date little experimental benchmarking data on the preferred metal binding sites, which is essential to understand the character of the metal-fullerene bonds, which in turn allows to better understand and/or improve the design of fullerene-based functional materials. Surprisingly, even for a single metal atom, there is no unambiguous experimental evidence regarding its binding site and interaction nature with the $C_{60}$ surface.[13] For example, employing X-ray and ultraviolet photoelectron spectroscopy, Kröger et al. showed that the nature of the interaction between Au and $C_{60}$ is covalent, but no signature of charge transfer was found in their study.[14] Lyon and Andrews concluded, based on matrix isolation infrared spectroscopy, that the Au atom preferentially binds on the pentagonal ring of $C_{60}$, but also pointed out that their conclusion may be subject to a matrix effect.[15] In contrast, the vertex binding site ($\eta^1$) has been found to be most stable for $C_{60}Au$, from DFT calculations including spin-orbit coupling.[16, 17] A third possible Au binding site was obtained in the work of Shukla et al., who found the bridge site when $C_{60}$ is sandwiched between two Au clusters.[11] Although the bonding scheme in sandwich-type complexes may be different from the $C_{60}$ examples with single metal atoms, these examples illustrate the challenge in achieving firm conclusions on the binding sites of metal atoms or clusters on the $C_{60}$ surface and thus their bonding nature. Using "indirect" structural information based on mass spectrometry, Kaya and co-workers suggested that the vanadium cation ($V^+$) binds above the hexagonal ring of $C_{60}$,[18] as predicted by several density functional theory (DFT) calculations,[13, 19] but in this case conclusive evidence is missing as well.

Recently, we have developed an experimental protocol to measure the laboratory infrared spectra of [$C_{60}$-Metal]$^+$ complexes via messenger-tagged infrared multiple photon dissociation (IRMPD) spectroscopy.[20-23] These fundamental experimental studies have an impact in multiple fields, including catalysis and astrophysics. We showed that a single vanadium cation supported on $C_{60}$, can efficiently catalyze the water splitting process to produce $H_2$ upon infrared light absorption, demonstrating the importance of carbonaceous supports in single atom catalysts,[20, 21] and proposed that the complexes of fullerene with cosmic abundant metals such as iron can be promising carriers of astronomical unidentified infrared bands.[23] However, fundamental questions regarding the exact binding site and detailed interaction nature of the metal with $C_{60}$ remain to be answered. In this work, we conclude from our combined IRMPD spectroscopy and comprehensive theoretical calculations that the most probable location of $V^+$ on $C_{60}$ is above the pentagon. We also explore the bonding nature and the origin of the stability of the fullerene-metal complex.

**Results and discussion**

The laboratory infrared spectrum of the gas-phase $C_{60}V^+$ complex was measured via IRMPD spectroscopy using the intense far- to mid-infrared light from the Free Electron Laser for Intra-Cavity Experiments (FELICE).[24] Recently, we reported the IRMPD spectrum of $C_{60}V^+$ using $D_2$ as a messenger in the spectral range of 400–1600 $cm^{-1}$ (see details in Methods),[23] but the presence of $D_2$ molecules leads to spectral features containing effects from the $D_2$ tagging with several small features around 500 and 750 $cm^{-1}$ due to $D_2$ motion.

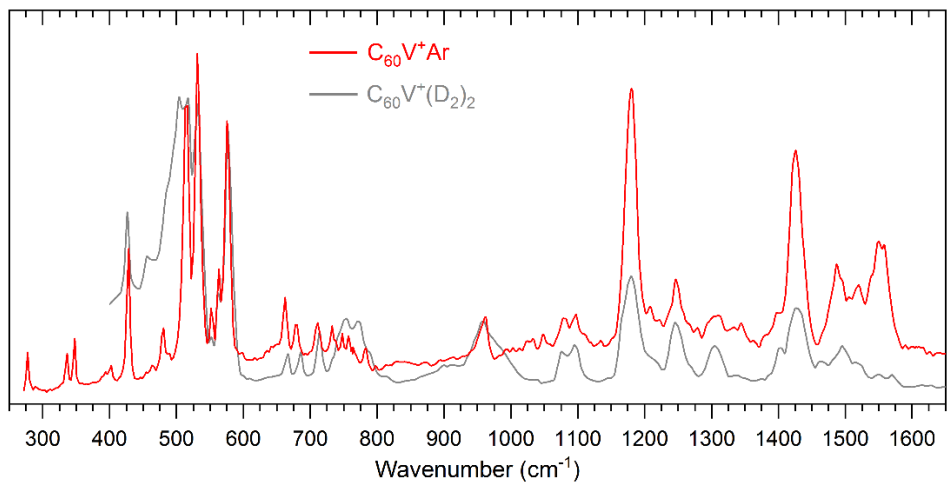

**Figure 1.** IRMPD spectrum of Ar-tagged $C_{60}V^+$ in 250–1700 $cm^{-1}$ range (red). The spectrum of $D_2$-tagged $C_{60}V^+$ with two tagging $D_2$ molecules in 400–1700 $cm^{-1}$ range (gray) is replotted from ref. **23** for comparison.

To obtain higher-quality IRMPD spectra of $C_{60}V^+$, we synthesized Ar-tagged $C_{60}V^+$ complexes. The calculated $Ar–C_{60}V^+$ binding energy of only 3.6 kJ/mol is much lower than the $D_2–C_{60}V^+$ binding energy of 30.5 kJ/mol. A lower messenger binding energy means it has less influence on both the geometric and the electronic properties of the parent species (Figures S1-S3).[25] Simultaneously, the photon energy of 300 $cm^{-1}$ light is about 3.6 kJ/mol, implying that fragmentation of the Ar-tagged $C_{60}V^+$ complex can occur by absorbing a single photon above 300 $cm^{-1}$ (see molecular dynamics simulations below). Therefore, the IRMPD spectrum of the Ar-tagged $C_{60}V^+$ complex should be close to a linear absorption spectrum.

**Figure 1** shows the infrared spectrum of Ar-tagged $C_{60}V^+$, and for comparison, the $D_2$-tagged $C_{60}V^+$ with two tagging $D_2$ molecules is replotted from ref. 23 as well. The comparison on the one hand shows that the infrared spectrum using an Ar-tag has higher spectral resolution as seen by the sharper features, presumably because fewer photons must be absorbed to fragment this complex as compared to the $D_2$-tagged $C_{60}V^+$. On the other hand, **Figure 1** shows that the two spectra are almost identical if disregarding the peak intensities and specific features in the 400–500 $cm^{-1}$ range and around 750 $cm^{-1}$, verifying our previously assigned effect of the $D_2$ tag on the vibrational spectrum of $C_{60}V^+$.[23] In addition, there are several new bands below 400 $cm^{-1}$, which could be characteristic for the $C_{60}$-metal interactions (see Figures S4-S5).

To extract information about the vanadium binding site, we first carried out quantum chemical calculations to obtain the candidate structures of the $C_{60}V^+$ complex. Because of the $I_h$ symmetry of $C_{60}$, only five different binding sites, i.e., $\eta^6$, $\eta^5$, $\eta^{2(6-6)}$, $\eta^{2(5-6)}$, and $\eta^1$ need to be considered.[20] In those structures, the V atom binds with a hexagonal center, a pentagonal center, a bridge of two hexagons, a bridge of a hexagon and a pentagon, and atop of a C atom, respectively. Since the ground state of $V^+$ is $[Ar]3d^34s^1$, the interaction of $V^+$ with the closed-shell molecule $C_{60}$ to form $C_{60}V^+$ could result in S = 2, 1, and 0 spin states, which all have been considered.

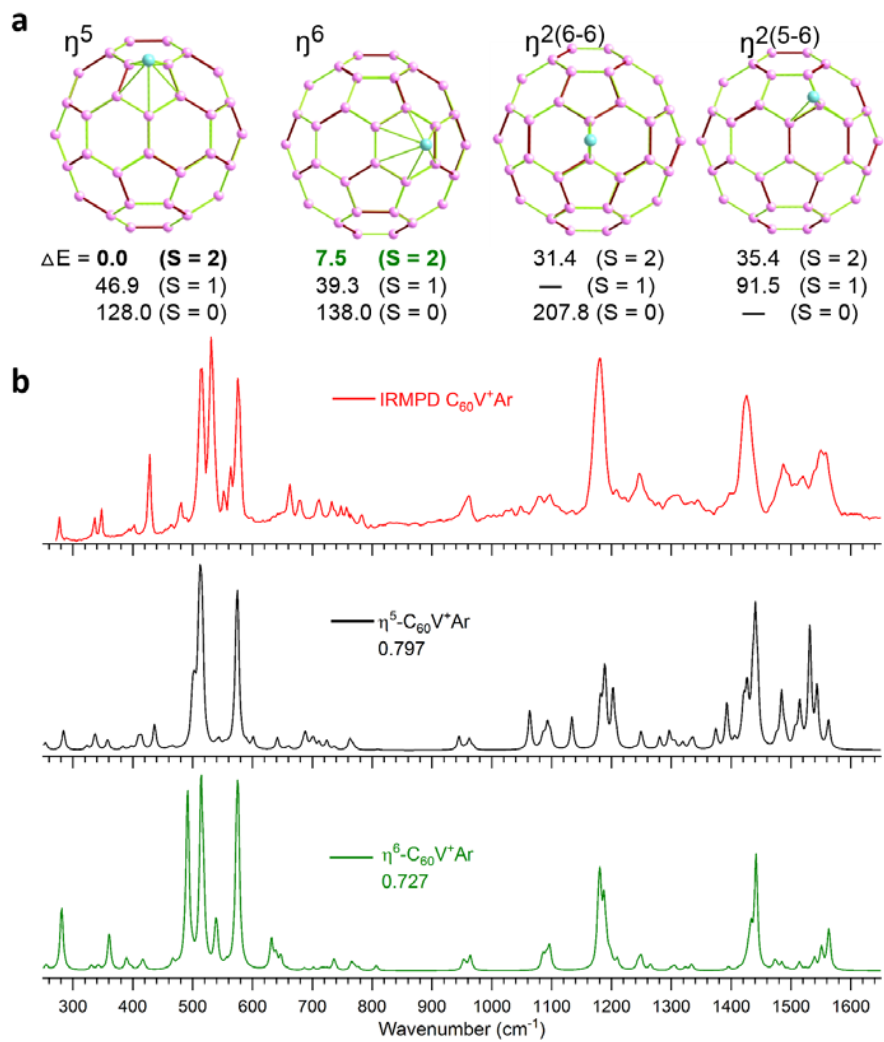

**Figure 2.** Experimental infrared spectrum of $C_{60}V^+Ar$ and its comparison with theoretical calculations. **a.** The four optimized structures, i.e., $\eta^5$, $\eta^6$, $\eta^{2(6-6)}$, and $\eta^{2(5-6)}$, and their relative stabilities ($\Delta E$, in kJ/mol) at BPW91/6-31G(d) level. **b.** Comparison of the experimental spectrum with simulated ones (broadened using Lorentzian line shapes of 6 cm$^{-1}$ full width at half maximum) for the lowest energy spin states (S = 2) in $\eta^5$ and $\eta^6$ configurations at the BPW91/6-31G(d) level of theory. The cosine similarity scores are indicated for the simulated spectra.

In **Figure 2a**, four optimized structures of $C_{60}V^+$, including $\eta^5$, $\eta^6$, $\eta^{2(6-6)}$, and $\eta^{2(5-6)}$, at BPW91/6-31G(d) level, together with their relative stabilities, are presented. The $\eta^1$ structure was unstable in the calculations and converged to either $\eta^5$ or $\eta^6$, and the calculations show that the high spin states have much higher stabilities, by over 30 kJ/mol, than the low spin states. Based on energetics, the two bridge structures, $\eta^{2(6-6)}$ and $\eta^{2(5-6)}$, can be ruled out as they are much higher in energy than the most stable $\eta^5$ structure by more than 30 kJ/mol. However, given that the typical accuracies of DFT methods are not better than 10 kJ/mol, the current energy difference between the $\eta^5$ and $\eta^6$ structures is insufficient to rule out either one, in addition to the possibility that higher energy isomers might be kinetically trapped due to large barriers (see below). Hence, both structures could coexist in the experiment. One could argue that the current bias towards the $\eta^5$ structure of 7.8 kJ/mol (Gibbs free energy at 300 K) lower in energy, which increases to 10.6 kJ/mol when going to the larger def2-TZVP basis set, strengthens the case for the pentagon, but these arguments remain tentative, if only because full thermalization in the experiment cannot be assumed.

Given the above energetics, a comparison of the experimental spectrum with the simulated ones is shown only for the $\eta^5$ or $\eta^6$ structures in S = 2 spin states in **Figure 2b** (a comparison with other isomers is presented in Figure S6). However, solely from a visual inspection, it is difficult to make a conclusive assignment of the experimental spectrum to any of the structures alone. For example, in the 900–1600 range, the match with $\eta^6$ appears quite good, apart from a missing band just below 1500 cm$^{-1}$. However, below 800 cm$^{-1}$, there are more discrepancies, e.g., few spectral features around 700 cm$^{-1}$, a relatively intense band at 630 cm$^{-1}$, and a clear mismatch below 450 cm$^{-1}$. In turn, the lowest energy $\eta^5$ structure overall does quite well, but fails to accurately predict the splitting of the 510–530 cm⁻¹ experimental band with a small frequency difference (see the vibrational vectors at 500 and 512 cm⁻¹ in Figure S4) in the predicted double-peak feature. Additionally, the relatively strong sharp experimental features below 500 cm⁻¹ (see close-up in the 250–650 cm⁻¹ spectral range in Figure S5 and the related vibrational mode vectors in Figure S4), computationally identified as motions directly involving the V–$C_{60}$ bond, do not allow for an unambiguous assignment.

Consequently, other analytical methods need to be considered. A useful tool to address this problem is the cosine similarity test.[26] This test, as detailed in the Methods section, was developed as a quantitative measure to evaluate the similarity between two curves. It has been successfully employed as an objective metric to evaluate the agreement between calculated and experimental infrared spectra.[27, 28] The calculated cosine similarity scores are provided in **Figures 2b** and **S6**; the η⁵ structure has the highest score among the four Ar-tagged structural isomers. We tested the application of uniform scaling factors for the calculated harmonic frequencies and the same trend for the calculated cosine similarity scores is obtained. While the value of 0.797 (a value of 1 should be obtained for identical spectra) reflects some disagreement between DFT calculations and experiments as illustrated in the previous section, it is about 0.07 larger than the value of 0.727 for the η⁶ $C_{60}V^+$. According to previous studies,[26-28] such difference allows to differentiate the η⁵ structure from the η⁶ structure, making the quantitative cosine similarity analysis objectively favoring the assignment of the experimental spectrum to η⁵ $C_{60}V^+$.

From the above energetic stability and spectral comparison analysis, we conclude that η⁵ $C_{60}V^+$ is the structure existing in our experiment. This contrasts to previous theoretical predictions from Sankar De, et al.[13] and Robledo et al.[19] that the η⁶ structure is most stable. Sankar De, et al employed the VASP program to optimize the different isomers of $C_{60}V^+$ and calculate their stabilities, but this method cannot describe the exact atomic configuration of V. The B3LYP functional used by Robledo et al. gives, according to our systematic calculations in Table S1, an almost degenerate stability for η⁵ and η⁶ $C_{60}V^+$ with the η⁵ structure being only 0.9 kJ/mol more stable. Such small energy difference does not allow a meaningful conclusion about the relative stability because the typical accuracies of DFT methods are not better than 10 kJ/mol.

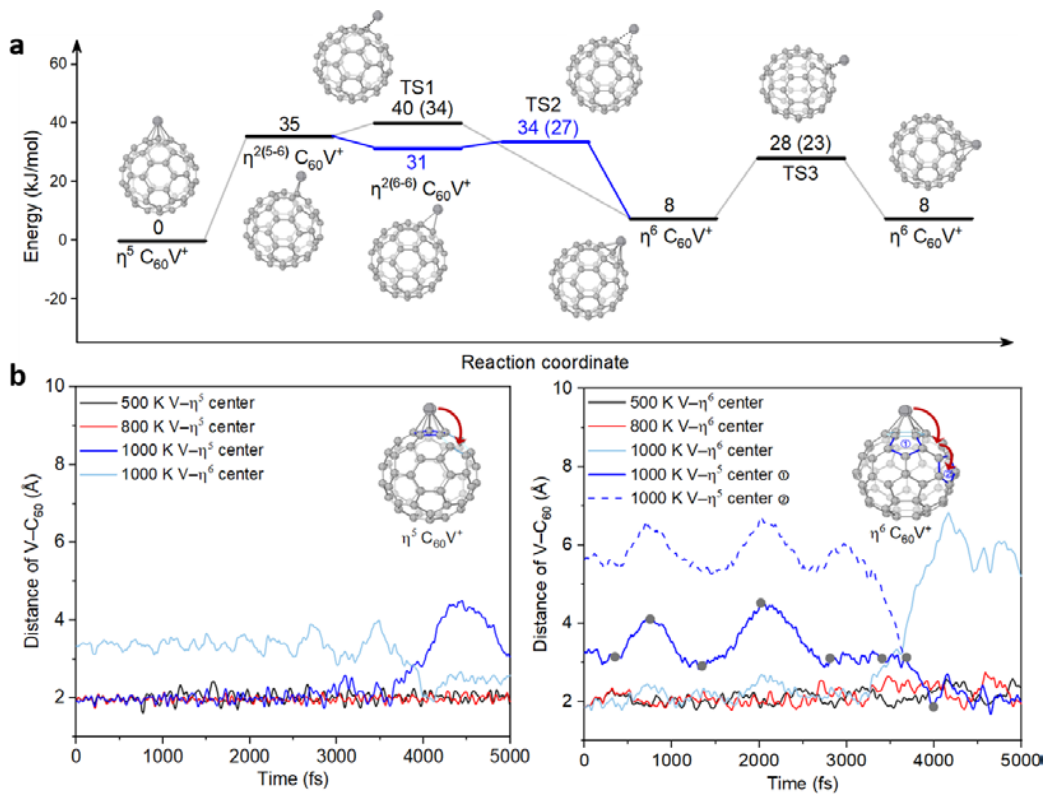

**Figure 3.** (a) Isomerization pathways linking the four $C_{60}V^+$ isomers calculated at BPW91/6-31G(d) level with all structures shown with a similar view. Two different pathways connecting the interconversion between η⁵ and η⁶ $C_{60}V^+$ are shown in black and blue, respectively. For transition states, the energies without ZPE correction are also presented in parentheses; (b) Distances between V and the center of the pentagon and hexagon of $C_{60}$ in 5 ps BOMD simulations at 500, 800, and 1000 K in η⁵ $C_{60}V^+$ (left panel) and η⁶ $C_{60}V^+$ (right panel), respectively. Red arrows are utilized to guide the eye about the conversion pathway during the simulation. In right panel, the distances between V and the center of the closest pentagon (center①) and the finally resultant pentagon (center②) in the simulation are shown respectively in solid and dashed lines. Snapshots indicated with grey circles are provided in Figure S7 to show the location changes of V during BOMD simulations.

In addition, Nagao et al. employed the 18-electron counting rule to explain the number of CO molecules that can bind with $C_{60}V^+$ in their mass spectrometric study, which suggests that $V^+$ binds with a hexagon of $C_{60}$ with a $\eta^6$ configuration.[8] That empirical interpretation does not allow to draw solid conclusion, emphasizing the critical importance of spectroscopic data. It should be noted that the calculated infrared spectra of $C_{60}V^+$ with and without Ar tag show essentially no difference as supported by the detailed comparison and calculated cosine similarity scores well above 0.9 (Figure S1). The minor effect of the Ar tag on the structure of $C_{60}V^+$ has been verified by calculating the root-mean-square deviation (RMSD) values below 0.01 Å between the $C_{60}V^+$ and $C_{60}V^+Ar$ geometries (Figure S2).

In the following, we seek to assess the thermal stability of $C_{60}V^+$ by exploring the isomerization pathways of its four isomers using both quantum chemical calculations and Born-Oppenheimer molecular dynamics (BOMD) simulations, as well as the bonding nature by conducting energy decomposition analysis (EDA). **Figure 3a** presents the isomerization pathways of the four $C_{60}V^+$ isomers calculated at BPW91/6-31G(d) level of theory. It can be seen that interconversion barriers are relatively large, although not completely insurmountable. The calculated pathways show that that the interconversion between $\eta^5$ and $\eta^6$ $C_{60}V^+$ can either go directly through $\eta^{2(5-6)}$ (black), or via both $\eta^{2(5-6)}$ and $\eta^{2(6-6)}$ (blue). Note, the isomerization between two bridge-site isomers, i.e., $\eta^{2(5-6)}$ and $\eta^{2(6-6)}$, should be via a $\eta^1$ structure; unfortunately, the transition state structure could not be located, implying that it might be a barrierless process.

**Figure 3b** shows the change of the $V-C_{60}$ distances in BOMD simulations of 5 ps at 500, 800, and 1000 K for both $\eta^5$ $C_{60}V^+$ (left panel) and $\eta^6$ $C_{60}V^+$ (right panel). At low temperatures of 500 and 800 K, the $V-C_{60}$ distances do not change significantly for both isomers, indicating no interconversion between the different isomers of $C_{60}V^+$. This is consistent with the relatively large interconversion barriers of around 30 kJ/mol between different isomers starting from either $\eta^5$ or $\eta^6$ $C_{60}V^+$. However, at an elevated temperature, the internal energy of $C_{60}V^+$ increases and its thermal motion makes it possible for the $V^+$ to reach the different binding sites. For both $\eta^5$ and $\eta^6$ $C_{60}V^+$ isomers without Ar tag, the simulations show that both $V-\eta^5$ and $V-\eta^6$ $C_{60}V^+$ distances suddenly increase significantly around 3.5 ps at 1000K; while when tagging with Ar, a significant increase in the $V-\eta^6$ $C_{60}V^+$ center distance is already observed at 800 K for the $\eta^6$ isomer. This distance remains relatively constant for the $\eta^5$ isomer during the 5 ps simulation time at 800 K (see Figure S8 that also includes the change of $V-Ar$ distances in Ar-tagged $C_{60}V^+$ structures). The interconversion starting from $\eta^6$ $C_{60}V^+$ is complicated, as it first starts converting to another $\eta^6$ $C_{60}V^+$ at about 3.5 ps and then to $\eta^5$ $C_{60}V^+$ at around 4.0 ps at a temperature of 1000 K (right panel in **Figure 3b**), in consistent with the calculated isomerization pathways (**Figure 3a**). It then maintains the $\eta^5$ $C_{60}V^+$ structure for the remaining simulation time, implying the slightly higher stability of $\eta^5$ $C_{60}V^+$ compared to $\eta^6$ $C_{60}V^+$. Besides, the simulations show that the Ar-tag gets detached from $C_{60}V^+$ in the temperature range of 500–800 K for both $\eta^5$ and $\eta^6$ $C_{60}V^+$ (Figure S8), in agreement with their low $Ar-C_{60}V^+$ binding energies of only 3–4 kJ/mol from DFT calculations.

To obtain a more direct view about the interconversion of the four $C_{60}V^+$ isomers, we further increased the simulation temperature to 2000 K, assuming that $C_{60}V^+$ may freely access its different isomers at such a high temperature. In Figure S9, we plotted the statistical trajectories of $C_{60}V^+$ in different isomers from a 5 ps BOMD simulation at 2000 K with $\eta^5$ $C_{60}V^+$ as the starting point. It can be seen that the four isomers indeed can convert to each other, and we summed the total time that each isomer stays to be 27%, 24%, 34%, and 15% for $\eta^5$, $\eta^6$, $\eta^{2(5-6)}$, and $\eta^{2(6-6)}$, respectively. Ideally, such values could represent the isomeric populations during the simulation and reflect the relative stability of the isomers. However, 5 ps simulation time may not be long enough to allow the four isomers to achieve a statistical thermodynamic equilibrium, under which condition the population of the isomers would follow a Boltzmann distribution depending on their thermodynamic stability. Movies showing the trajectories of the BOMD simulations are provided in supporting information.

**Table 1.** Bond dissociation energies $(D_o)$[a] and results from the energy decomposition analysis (EDA-NOCV)[b] for four $C_{60}V^+$ isomers. All energies are in kJ/mol.

| | $\eta^5$ $C_{60}V^+$ | $\eta^6$ $C_{60}V^+$ | $\eta^{2(6-6)}$ $C_{60}V^+$ | $\eta^{2(5-6)}$ $C_{60}V^+$ |
|---|---|---|---|---|
| $D_o$ | -264.3 | -261.1 | -238.1 | -234.3 |
| $\Delta E_{int}$ | -266.9 | -258.8 | -259.7 | -260.8 |
| $\Delta E_{pauli}$ | 472.1 | 446.6 | 331.1 | 317.7 |
| $\Delta E_{disp}$[c] | -25.4 (3.4%) | -25.3 (3.6%) | -31.7 (5.4%) | -30.5 (5.3%) |
| $\Delta E_{elstat}$[c] | -199.4 (27.0%) | -193.5 (27.4%) | -181.0 (30.6%) | -164.2 (28.4%) |
| $\Delta E_{orb}$[c] | -514.2 (69.6%) | -486.6 (69.0%) | -378.1 (64.0%) | -383.8 (66.3%) |
| $\Delta E_{orb(1)}$[d] | -150.9 (29.3%)<br>$V^+ (d_{zx}) \leftarrow C_{60}$ | -102.8 (21.1%)<br>$V^+ (d_{zx}) \leftarrow C_{60}$ | -151.8 (40.1%)<br>$V^+ (d_{zx}) \rightarrow C_{60}$ | -138.6 (36.1%)<br>$V^+ (d_{zx}) \rightarrow C_{60}$ |
| $\Delta E_{orb(2)}$[d] | -87.9 (17.1%)<br>$V^+ (d_{yz}) \rightarrow C_{60}$ | -99.8 (20.5%)<br>$V^+ (d_{x2-y2}) \rightarrow C_{60}$ | -114.8 (30.4%)<br>$V^+ (d_{z2}) \leftarrow C_{60}$ | -106.4 (27.7%)<br>$V^+ (d_{z2}) \leftarrow C_{60}$ |
| $\Delta E_{orb(3)}$[d] | -84.3 (16.4%)<br>$V^+ (d_{x2-y2}) \rightarrow C_{60}$ | -67.0 (13.8%)<br>$V^+ (d_{xy}) \rightarrow C_{60}$ | -23.4 (6.2%)<br>$V^+$ polarization | -43.7 (11.4%)<br>$V^+$ polarization |
| $\Delta E_{orb(4)}$[d] | -52.6 (10.2%)<br>$V^+ (d_{xy}) \rightarrow C_{60}$ | -56.5 (11.6%)<br>$V^+ (d_{yz}) \rightarrow C_{60}$ | -17.5 (4.6%)<br>$V^+ (d_{yz}) \rightarrow C_{60}$ | -18.0 (4.7%)<br>$V^+ (d_{yz}) \rightarrow C_{60}$ |
| $\Delta E_{orb(5)}$[d] | -41.5 (8.1%)<br>$V^+ (d_{z2}) \rightarrow C_{60}$ | -33.9 (7.0%)<br>$V^+ (d_{z2}) \rightarrow C_{60}$ | -9.4 (2.5%)<br>$V^+ (d_{xy}) \rightarrow C_{60}$ | -16.4 (4.3%)<br>$V^+ (d_{xy}) \rightarrow C_{60}$ |
| $\Delta E_{orb(rest)}$[d] | -97.0 (18.9%) | -126.6 (26.0%) | -61.2 (16.2%) | -60.7 (15.8%) |

[a]Values calculated at the BPW91/6-31G(d) level; [b]Values calculated at the BP86-D3(BJ)/TZ2P//BPW91/6-31G(d) level using $V^+$ (quintet) and $C_{60}$ (singlet) fragments; [c]Values in parentheses are the relative contributions to the total attractive interactions $\Delta E_{disp} + \Delta E_{elstat} + \Delta E_{orb}$; [d]Values in parentheses are the relative contributions to $\Delta E_{orb}$.

The bonding nature between $V^+$ and $C_{60}$ is investigated using energy decomposition analysis combined with natural orbitals for chemical valence (EDA-NOCV) at the BP86-D3(BJ)/TZ2P level for all the four $C_{60}V^+$ isomers optimized at BPW91/6-31G(d) level.[29] **Table 1** summarizes the calculated results for the interactions between the $V^+$ (quintet) and $C_{60}$ (singlet) fragments. The calculated interaction energies $\Delta E_{int}$ are in good agreement with the calculated bond dissociation energies ($D_o$). The main contributions to the total bonding interactions $\Delta E_{int}$ come from the electrostatic term $\Delta E_{elstat}$ and the orbital interaction term $\Delta E_{orb}$. In particular, both $\Delta E_{orb}$ and $\Delta E_{elstat}$ have the highest absolute values for $\eta^5$ $C_{60}V^+$. The larger bonding interactions of $\eta^5$ $C_{60}V^+$ compared to those of the other three isomers is consistent with the higher stability of $\eta^5$ $C_{60}V^+$ found at the BPW91/6-31G(d) level. Further decomposition of the total orbital interactions $\Delta E_{orb}$ results in the individual orbital terms including $\Delta E_{orb(1)}-\Delta E_{orb(5)}$, corresponding to the five d orbitals of V. This decomposition analysis shows subtle changes of the orbital orders among the four isomers, suggesting a different bonding scenario.

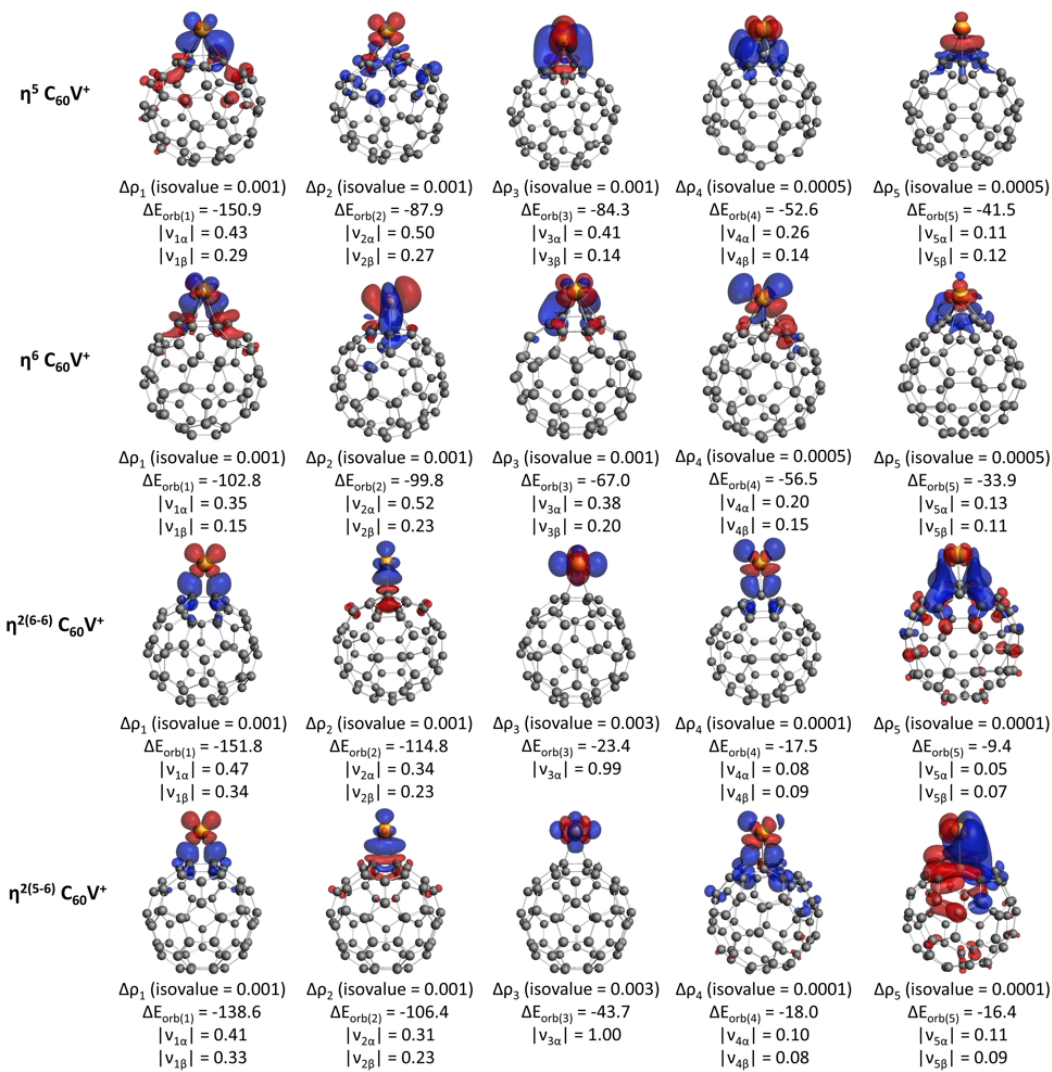

| | | | | |
|---|---|---|---|---|
| $\eta^5$ $C_{60}V^+$ | | | | |
| $\Delta\rho_1$ (isovalue = 0.001) | $\Delta\rho_2$ (isovalue = 0.001) | $\Delta\rho_3$ (isovalue = 0.001) | $\Delta\rho_4$ (isovalue = 0.0005) | $\Delta\rho_5$ (isovalue = 0.0005) |
| $\Delta E_{orb(1)}$ = -150.9 | $\Delta E_{orb(2)}$ = -87.9 | $\Delta E_{orb(3)}$ = -84.3 | $\Delta E_{orb(4)}$ = -52.6 | $\Delta E_{orb(5)}$ = -41.5 |
| $|v_{1a}|$ = 0.43 | $|v_{2a}|$ = 0.50 | $|v_{3a}|$ = 0.41 | $|v_{4a}|$ = 0.26 | $|v_{5a}|$ = 0.11 |
| $|v_{1\beta}|$ = 0.29 | $|v_{2\beta}|$ = 0.27 | $|v_{3\beta}|$ = 0.14 | $|v_{4\beta}|$ = 0.14 | $|v_{5\beta}|$ = 0.12 |

| | | | | |
|---|---|---|---|---|
| $\eta^6 V^+$ | | | | |
| $\Delta\rho_1$ (isovalue = 0.001) | $\Delta\rho_2$ (isovalue = 0.001) | $\Delta\rho_3$ (isovalue = 0.001) | $\Delta\rho_4$ (isovalue = 0.0005) | $\Delta\rho_5$ (isovalue = 0.0005) |
| $\Delta E_{orb(1)}$ = -102.8 | $\Delta E_{orb(2)}$ = -99.8 | $\Delta E_{orb(3)}$ = -67.0 | $\Delta E_{orb(4)}$ = -56.5 | $\Delta E_{orb(5)}$ = -33.9 |
| $|v_{1a}|$ = 0.35 | $|v_{2a}|$ = 0.52 | $|v_{3a}|$ = 0.38 | $|v_{4a}|$ = 0.20 | $|v_{5a}|$ = 0.13 |
| $|v_{1\beta}|$ = 0.15 | $|v_{2\beta}|$ = 0.23 | $|v_{3\beta}|$ = 0.20 | $|v_{4\beta}|$ = 0.15 | $|v_{5\beta}|$ = 0.11 |

| | | | | |
|---|---|---|---|---|
| $\eta^{2(6-6)}$ $C_{60}V^+$ | | | | |
| $\Delta\rho_1$ (isovalue = 0.001) | $\Delta\rho_2$ (isovalue = 0.001) | $\Delta\rho_3$ (isovalue = 0.003) | $\Delta\rho_4$ (isovalue = 0.0001) | $\Delta\rho_5$ (isovalue = 0.0001) |
| $\Delta E_{orb(1)}$ = -151.8 | $\Delta E_{orb(2)}$ = -114.8 | $\Delta E_{orb(3)}$ = -23.4 | $\Delta E_{orb(4)}$ = -17.5 | $\Delta E_{orb(5)}$ = -9.4 |
| $|v_{1a}|$ = 0.47 | $|v_{2a}|$ = 0.34 | $|v_{3a}|$ = 0.99 | $|v_{4a}|$ = 0.08 | $|v_{5a}|$ = 0.05 |
| $|v_{1\beta}|$ = 0.34 | $|v_{2\beta}|$ = 0.23 | | $|v_{4\beta}|$ = 0.09 | $|v_{5\beta}|$ = 0.07 |

| | | | | |
|---|---|---|---|---|
| $\eta^{2(5-6)}$ $C_{60}V^+$ | | | | |
| $\Delta\rho_1$ (isovalue = 0.001) | $\Delta\rho_2$ (isovalue = 0.001) | $\Delta\rho_3$ (isovalue = 0.003) | $\Delta\rho_4$ (isovalue = 0.0001) | $\Delta\rho_5$ (isovalue = 0.0001) |
| $\Delta E_{orb(1)}$ = -138.6 | $\Delta E_{orb(2)}$ = -106.4 | $\Delta E_{orb(3)}$ = -43.7 | $\Delta E_{orb(4)}$ = -18.0 | $\Delta E_{orb(5)}$ = -16.4 |
| $|v_{1a}|$ = 0.41 | $|v_{2a}|$ = 0.31 | $|v_{3a}|$ = 1.00 | $|v_{4a}|$ = 0.10 | $|v_{5a}|$ = 0.11 |
| $|v_{1\beta}|$ = 0.33 | $|v_{2\beta}|$ = 0.23 | | $|v_{4\beta}|$ = 0.08 | $|v_{5\beta}|$ = 0.09 |

**Figure 4.** Plots of the deformation densities $\Delta\rho_{(i)}$ to $\Delta\rho_{(5)}$ associated to the pairwise interactions $\Delta E_{orb(1)}$ to $\Delta E_{orb(5)}$ in $C_{60}V^+$ with corresponding energies (kJ/mol) and eigenvalues ($e$) by using $V^+$ (quintet) and $C_{60}$ (singlet) as interacting fragments. The charge flow is from red to blue.

The effect of orbital interactions on the charge distribution can be seen from the shape of the deformation densities ($\Delta\rho$) associated with the orbital interactions. **Figure 4** shows the contour plots of the deformation densities $\Delta\rho_{(i)}$ to $\Delta\rho_{(5)}$ that are connected to the pairwise interactions $\Delta E_{orb(i)}$ to $\Delta E_{orb(5)}$ in $C_{60}V^+$ (**Table 1**). Among these components, charge transfer (donor-acceptor interactions between the two fragments) and polarization (unoccupied-occupied orbital mixing on $V^+$ due to the presence of $C_{60}$) are dominant contributions. In the case of $\eta^5$ $C_{60}V^+$, two different bonding types are revealed. First, the dative bond ($\Delta\rho_{(i)}$) is constructed by the donation from the highest occupied molecular orbital (HOMO) of $C_{60}$ to the $V^+$ $d_{zx}$ orbital, which contributes most to the total orbital interactions. Second, $\Delta\rho_{(2)}$–$\Delta\rho_{(5)}$ of $\eta^5$ $C_{60}V^+$ corresponds to the dative bond constructed by the donation from $d_{yz}$, $d_{x2-y2}$, $d_{xy}$, and $d_{z2}$ of $V^+$ to the lowest unoccupied molecular orbital (LUMO) of $C_{60}$, thus forming the four singly occupied orbitals of $C_{60}V^+$. The analysis suggests that orbital interactions play a crucial role in the charge transfer between $V^+$ and $C_{60}$. It is similar for the other three isomers, but delicate differences exist regarding the orbital energy orders and thus the electronic structures. Such delicate differences could lead to subtle variations in functionality of decorated fullerenes, and would cause $C_{60}$-metal complexes to have unique spectral signatures in the ultraviolet and optical wavelength region. That may also potentially be important to investigate their relevance to the astronomical mysterious diffuse interstellar bands.[23, 30, 31]

## Conclusion

In summary, we investigated the binding position of a single $V^+$ on the buckyball surface by combining Ar-tagging infrared spectroscopy, quantum chemical calculations, and molecular dynamics simulations. Comparison between the experimental and calculated infrared spectra with the aid of an objective similarity analysis suggests that Ar-tagging has a negligible effect on both the geometric and vibrational structures of $C_{60}V^+$, and that the most probable binding site for $V^+$ is the center of a $C_{60}$ pentagon ($\eta^5$ binding). Calculated isomerization pathways and Born-Oppenheimer molecular dynamics simulations at a range of temperatures further show a high thermal stability of $C_{60}V^+$ beyond temperatures of 800 K. Chemical bonding analysis reveals that the large orbital and electrostatic interactions are responsible for the high stability of $C_{60}V^+$, in particular for the $\eta^5$ structure. This information would be of critical importance for the rational design of fullerene-based functional materials.

## Methods

**Experimental.** The $C_{60}V^+$ complex is synthesized in vacuum by a dual-target dual-laser ablation source, which is a modified version of the Smalley-type laser vaporization source used in the discovery of $C_{60}$ in 1985 (ref.[32]). Our cluster source has been described in detail previously.[33] For the synthesis of $C_{60}V^+$ we use a bulk vanadium target and a fullerene target that is obtained by cold-pressing $C_{60}$ powder.[20-23, 34] Both targets are vaporized by 532 nm laser pulses from two independent Nd:YAG lasers, both operated at 10 Hz repetition rates. Note that the laser for the fullerene target is off focused on the fullerene target and is operated with much lower power compared to the laser for the vanadium target to avoid fragmentation of the fullerenes. The vaporized neutral $C_{60}$ molecules and metal plasma, containing metal cations, collide with each other in the presence of He gas, introduced through a pulsed valve with a 6 bar stagnation pressure, which triggers formation and cooling of the complexes. We assume [$C_{60}$-Metal]$^+$ complexes to be thermalized to room temperature before expansion into vacuum, moderately cooling their internal degrees of freedom.

The high stability of $C_{60}V^+$, characterized by a calculated binding energy of 2.82 eV at BPW91/6-31G(d) (see below) between $C_{60}$ and $V^+$, precludes the recording of high-quality infrared spectrum due to the inefficient dissociation induced by the infrared photon absorption as shown in our recent work.[23] Hence, the messenger tagging technique must be utilized, which has been well established in the past decades to obtain the vibrational or electronic spectra of molecular species via photodissociation.[35-37] Previously, we employed $D_2$ as tag for $C_{60}V^+$, but the calculated binding energy of $D_2$ with $C_{60}V^+$ is about 30.5 kJ/mol. In this work, we utilize Ar, which with a calculated binding energy of about 3.6 kJ/mol, corresponding to only one photon at 300 cm⁻¹, is a more sensitive messenger and simultaneously, through its weaker interaction, will lead to less changes to the intrinsic vibrational structure of the parent species. The Ar-tagged $C_{60}V^+$ complex was formed by seeding about 2% Ar into the He carrier gas. After expansion into vacuum through a conical nozzle, the cluster beam is formed and shaped by a 2 mm diameter skimmer and a 2 mm slit aperture, before entering into the extraction zone of a perpendicular reflectron time-of-flight (TOF) mass spectrometer.

IRMPD experiments are performed by overlapping the shaped cluster beam with IR light of the Free Electron Laser for Intra-Cavity Experiments FELICE.[24] The measurements were conducted with two FEL settings, covering the 250–750 and 600–1800 cm⁻¹ spectral ranges, at a repetition rate of 5 Hz. This allowed to alternatingly record mass spectra with and without IR laser interaction. Laser excitation in resonance with a vibrational mode heats up the clusters by infrared photon absorption and intra-molecular vibrational redistribution (IVR). When the internal energy of the cluster is high enough, fragmentation (dissociation) takes place on the time scale of the experiment via the lowest-energy fragmentation channel, in the current case desorption of Ar from $C_{60}V^+ \cdot Ar$. Here, the Ar atom serve as non-interfering weakly bound messenger that will be shed if IR radiation is resonantly absorbed, representing a sensitive probe by mass spectrometry. Care has been taken with the FEL light focusing to minimize saturation effects and to avoid destruction of the $C_{60}$ cage, which is confirmed by the absence of $C_{60-2m}^+$ fragments in the mass spectra. In the current experiment, absorption of one infrared photon might be enough to induce fragmentation due to the weak Ar–$C_{60}V^+$ interaction, making the experiment likely to be IRPD.

The FEL was scanned in wavenumber steps of 3 cm⁻¹, and the IR wavelength is calibrated using a grating spectrometer. The experiments allow to obtain IR spectra by comparing the mass spectrometric intensities of $C_{60}V^+$ and its Ar-tagged complex with [$I(v)$] and without ($I_0$) FEL light irradiation. We first calculate the branching ratio $B$ of the number of $C_{60}V^+ Ar_1$ ions to all $C_{60}V^+ Ar_{0,1}$ ions,

$$B = I[C_{60}V^+Ar_1] / \sum I[C_{60}V^+Ar_{0,1}]. \qquad (1)$$

Under the assumption of constant Ar adsorption rate, this eliminates fluctuations in the metal-fullerene synthesis. We then calculate the depletion $D(v)$ as a function of IR frequency $v$ by taking the natural logarithmic ratio of the branching ratios with and without IR irradiation,

$$D(v) = -\ln[B(v)/B_0]. \qquad (2)$$

The depletion $D(v)$ is divided by the laser pulse energy $E(v)$ to account for variation of the laser power and to approximate the infrared absorption cross section,

$$\sigma(v) = D(v)/E(v). \qquad (3)$$

The FEL laser pulse energy is reconstructed by measuring the pulse energy of a fraction of the pulse that is outcoupled of the FELICE cavity.[23] Typical pulse energies range from 1000 mJ at 400 cm⁻¹ to 200 mJ at 1600 cm⁻¹.

**Theoretical.** Five structures of $C_{60}V^+$, i.e., $\eta^5$, $\eta^6$, $\eta^{2(6-6)}$, $\eta^{2(6-5)}$, and $\eta^1$ have been fully optimized to obtain the most probable binding site of the vanadium ion on the $C_{60}$ surface. Endohedral structures are not considered as the experiments start with preformed $C_{60}$, while it is known that the activation energy for a metal atom entering into the $C_{60}$ cage is significant[38]. Six functionals, i.e., BPW91, PBE, PBE0, B3LYP, M06-2X, and ωB97XD, have been employed to justify the reliability of density functional theory (DFT) calculations. All these functionals calculate the structure in S = 2 spin state to be most stable (Table S1). From the comparison in Figure S10, it can be seen that BPW91 gives highest cosine similarity score in comparison with the experiment without applying any scaling factors, consistent with our previous findings on the good performance of BPW91 in simulating the infrared spectra of $C_{60}V^+(H_2O)_{1,2}$ complexes.[20, 21] Due to the large size of the studied complex, double-ζ quality basis sets (6-31G(d) and/or def2-SVP, see details in texts) were used for all calculations. Triple-ζ quality basis sets (6-311G(d), def2-TZVP, and Lanl2TZ) were also checked, showing consistent results with the double-ζ basis sets (Figure S11). The harmonic approximation was employed to perform the vibrational analysis of the calculated structures. The vibrational analysis is used on one hand to confirm that the calculated structures are real minima for ground states or have only one imaginary frequency for transition states, and on the other hand to simulate theoretical IR spectra to compare with the measured IR(M)PD spectra. All the calculations were conducted with Gaussian09 program package[39]. Energy decomposition analysis (EDA),[40] performed with the ADF 2020 software package,[41] was carried out at the BP86+D3(BJ)/TZ2P level to partition the bonding interaction energy $\Delta E_{int}$ between $V^+$ and $C_{60}$ into orbital ($\Delta E_{orb}$) and electrostatic ($\Delta E_{elstat}$) interactions, Pauli repulsion ($\Delta E_{Pauli}$), and dispersion ($\Delta E_{disp}$). The natural orbitals for chemical valence (NOCV) can decompose the orbital interaction term $\Delta E_{orb}$ into pairwise contributions, and plots of deformation densities were used to visualize the electron transfer caused by orbital interaction.[42, 43]

Born-Oppenheimer Molecular Dynamics (BOMD) simulations were carried out under canonical ensemble (NVT) using the PBE functional with the D3 dispersion correction[44] as implemented in the CP2K code.[45] The DZVP-MOLOPT-SR-GTH basis set was used in conjunction with Geodecker, Teter, and Hutter (GTH) pseudopotentials[46] with a plane wave cutoff energy of 300 Ry and a REL_CUTOFF of 40 Ry. The three orthorhombic lattice constants were set as 25 Å to avoid lateral interactions between the periodic images. The total simulation time was 5 ps with 1 fs per step.

**Cosine Similarity Score.** To quantitatively assess the agreement between the calculated and experimental infrared spectra, we employed the recently proposed cosine similarity score.[26-28] In this method the similarity is quantified as the cosine of the angle $\theta$ between two $n$-dimensional vectors, calculated using their normalized Euclidean dot product:

$$\cos(\theta) = \frac{A \cdot B}{||A|| \, ||B||} = \frac{\sum_{i=1}^n A_i B_i}{\sqrt{\sum_{i=1}^n A_i^2} \cdot \sqrt{\sum_{i=1}^n B_i^2}}$$

where A and B are two $n$-dimensional vectors with elements $A_i$ and $B_i$. This method assesses the degree to which the two vectors, representing the experimental and theoretical spectra in this case, are parallel. The outcomes range from 0 to 1, where a cosine value closer to 1 indicates a higher similarity. The intensity values in the computed spectrum are evaluated at the exact wavenumbers of the experimental spectrum, so that the two spectra have a common x-axis. To take into account possible anharmonic corrections for the calculated harmonic frequencies, the calculated spectra can be scaled before calculating the cosine similarity scores.

Kempkes et al.[27] proposed a slightly modified version of the above formula to make the cosine similarity scores more sensitive to the frequency overlap between bands in $A$ and $B$, and less to the deviations in their peak intensities. That may partially eliminate the potential effects of tagging and/or multiple photon absorption on the measured intensities. Both the experimental and calculated spectra are scaled to a maximum intensity of 1 and then the logarithm of these scaled values is taken as

$$A_i^{rev} = \log(\frac{A_i}{A_{max}} + c)$$

where $c$ is a constant that is identical for vectors $A$ and $B$. The value of $c$ is a compromise between being sensitive to low-intensity bands in the spectrum on the one hand and avoiding experimental noise affecting the similarity on the other hand. We followed this approach in the current work and used a $c$ value of 0.71 by testing a small set of experimental and computational spectra to give the best results.

**Supporting Information.** Additional theoretical results are available in the supporting information.

**Data availability.** The data that support the findings of this study are available from the corresponding author on reasonable request.

**Acknowledgements.** This work was supported by National Natural Science Foundation of China (92261101), the Innovation Capability Support Program of Shaanxi Province (2023-CX-TD-49), and the KU Leuven Research Council (project C14/22/103 and KA/20/045). G.-L.H acknowledge the support from the "Young Talent Support Plan" of Xi'an Jiaotong University, the "Selective Funding Project for Overseas Student Scientific and Technological Activities" of Shaanxi province, as well as the Fundamental

Research Funds for Central Universities, China. Part of the computational resources and services used in this work were provided by the VSC (Flemish Supercomputer Center), funded by the Research Foundation–Flanders (FWO) and the Flemish Government– department EWI. We gratefully acknowledge the Nederlandse Organisatie voor Wetenschappelijk Onderzoek (NWO) for the support of the FELIX Laboratory and thank the FELIX staff. G.-L.H gratefully acknowledges Prof. Su-Yuan Xie from Xiamen University for inspiring and encouraging discussions during the preparation of this manuscript.

**Author contributions**. G.-L.H conceived and coordinated the work. G.-L.H performed the IRMPD experiments with O.V.L., J.M.B., and E.J. J.X. and G.-L.H. conducted the theoretical calculations. G.L.H., E.J., and P.L. obtained the funding for this research. G.-L.H. wrote the manuscript with comments and suggestions from all authors. All authors approved the final version of the manuscript.

**Competing interests** The authors declare no competing interests.

# Supporting information

## Pentagon, Hexagon, or Bridge? Identifying the Location of a Single Vanadium Cation on Buckminsterfullerene Surface

Jianzhi Xu[1], Joost M. Bakker[2], Olga V. Lushchikova[2], Peter Lievens[3], Ewald Janssens[3*], Gao-Lei Hou[1*]

1. MOE Key Laboratory for Non-Equilibrium Synthesis and Modulation of Condensed Matter, School of Physics, Xi´an Jiaotong University, Xi´an, 710049 Shaanxi, China
2. FELIX Laboratory, Radboud University, Toernooiveld 7, 6525 ED Nijmegen, The Netherlands
3. Quantum Solid-State Physics, KU Leuven, Celestijnenlaan 200D, 3001 Leuven, Belgium

*Email: gaolei.hou@xjtu.edu.cn (G.-L.H.) ; ewald.janssens@kuleuven.be (E.J)

## Table of Contents

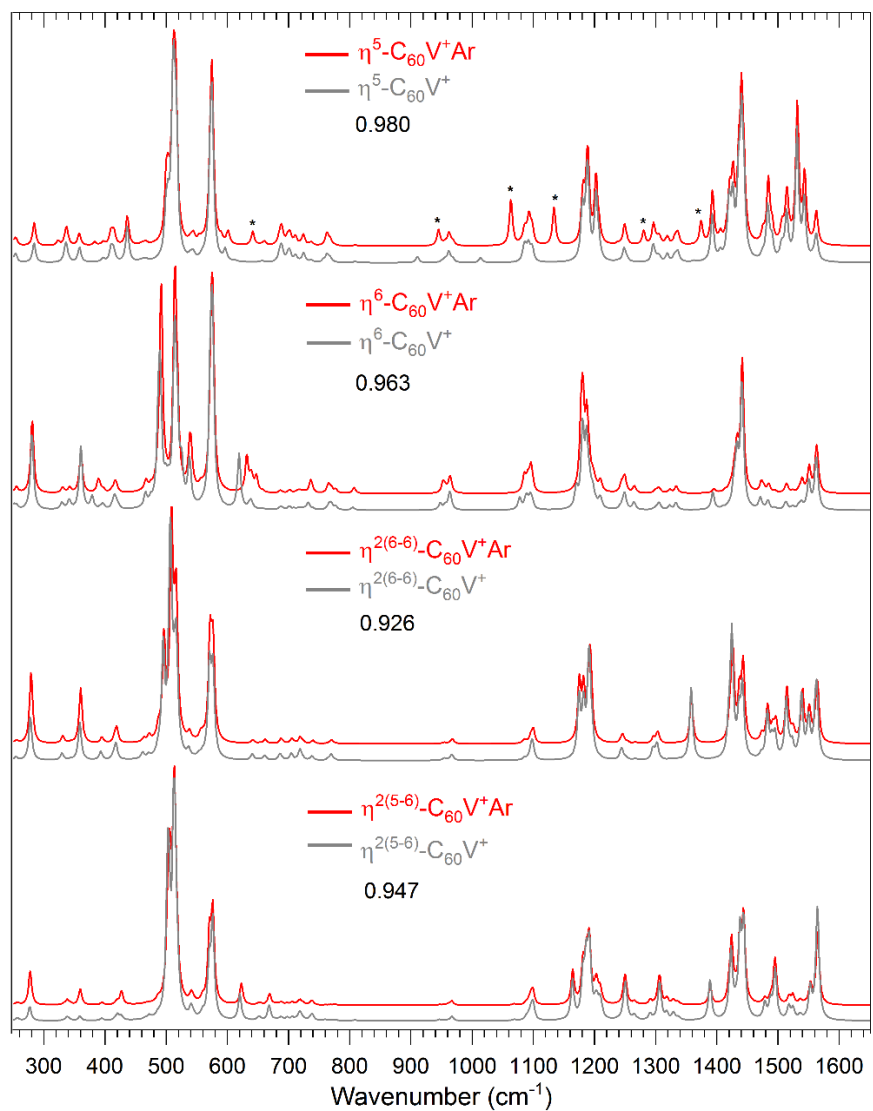

**Figure S1.** Simulated infrared spectra of $C_{60}V^+$ with and without Ar-tag at the BPW91/6-31G(d) level. The spectra are broadened using Lorentzian line shapes of 6 $cm^{-1}$ full width at half maximum. All four isomers have the S = 2 spin state. Several features marked with asterisks for the $\eta^5$ structure are from the motions of carbon cage activated by attaching Ar (Figure S3), and the other features are basically unaffected by Ar-tagging, supported by the calculated Cosine similarity scores for each isomer. Note that our calculations show that the Ar-involved vibrational modes appear below 200 $cm^{-1}$ (not shown on the figures).

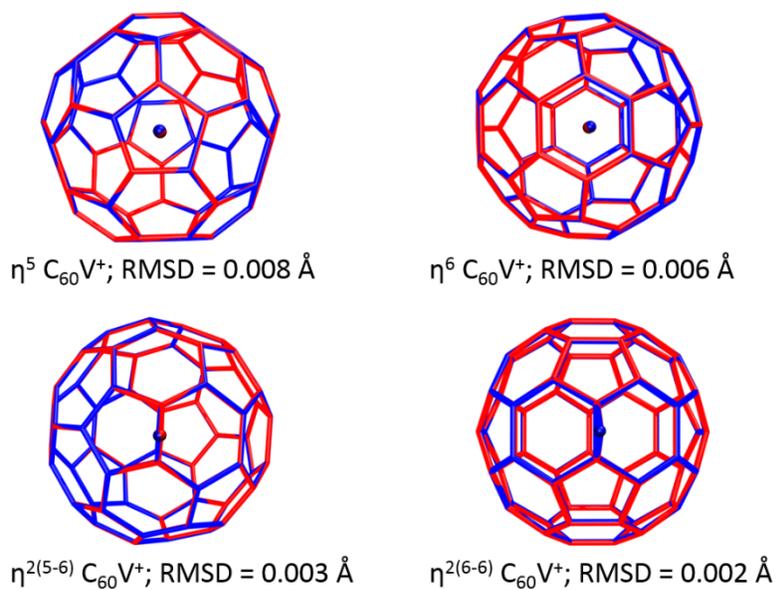

$\eta^5$ $C_{60}V^+$; RMSD = 0.008 Å     $\eta^6$ $C_{60}V^+$; RMSD = 0.006 Å

$\eta^{2(5-6)}$ $C_{60}V^+$; RMSD = 0.003 Å     $\eta^{2(6-6)}$ $C_{60}V^+$; RMSD = 0.002 Å

**Figure S2.** Overlaid structures and root-mean-square deviation (RMSD) for distances related to all atom positions for the four isomers of $C_{60}V^+$ with (in blue) and without Ar (in red) obtained at the BPW91/6-31G* level.

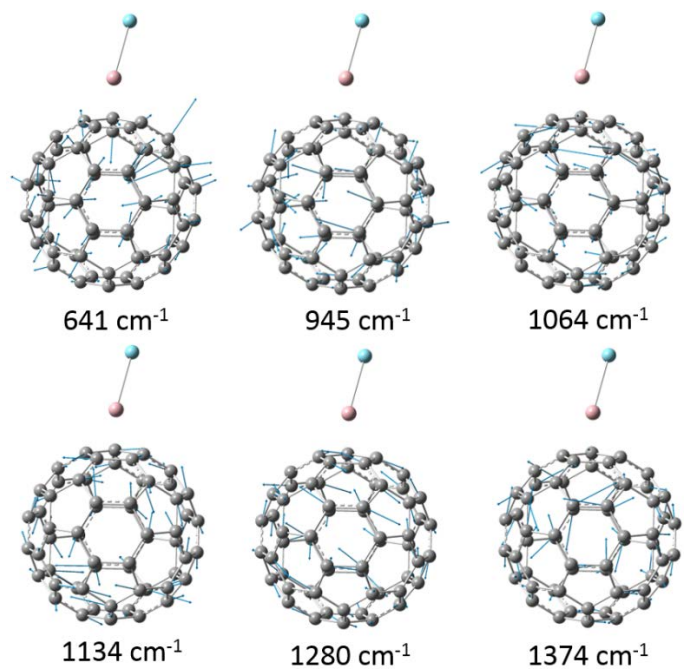

**Figure S3.** The vectors of the vibrational modes of $\eta^5\,C_{60}V^+$ that appear after Ar-tagging, but do not involve the motions of the Ar.

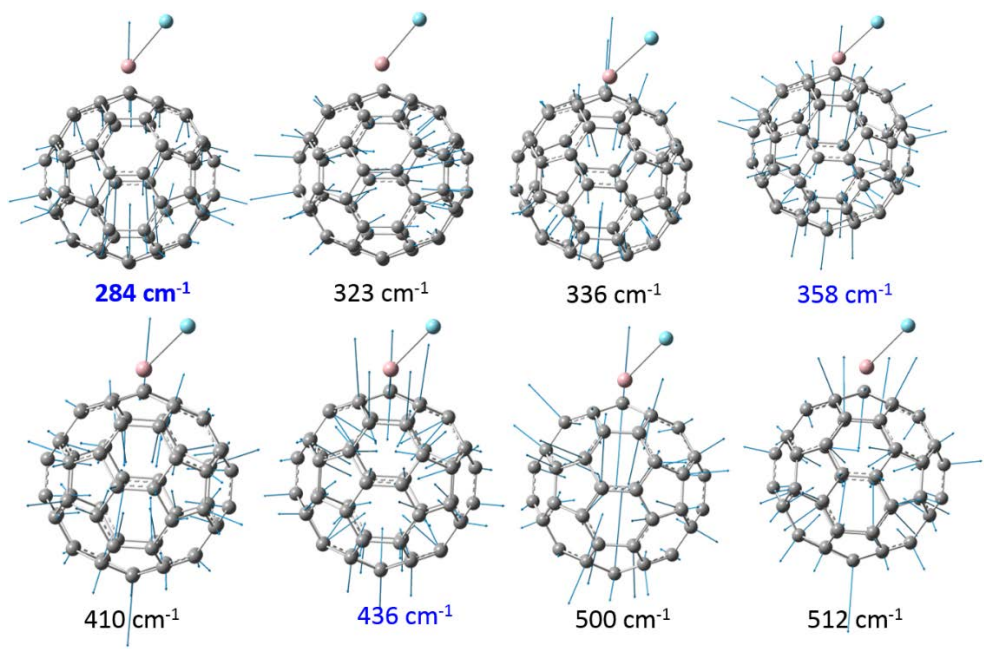

**Figure S4.** The vectors of selected vibrational modes of η$^5$ C$_{60}$V$^+$ in the 250–650 cm$^{-1}$ spectral range. The frequencies of vibrational modes that involve V motion are highlighted in blue.

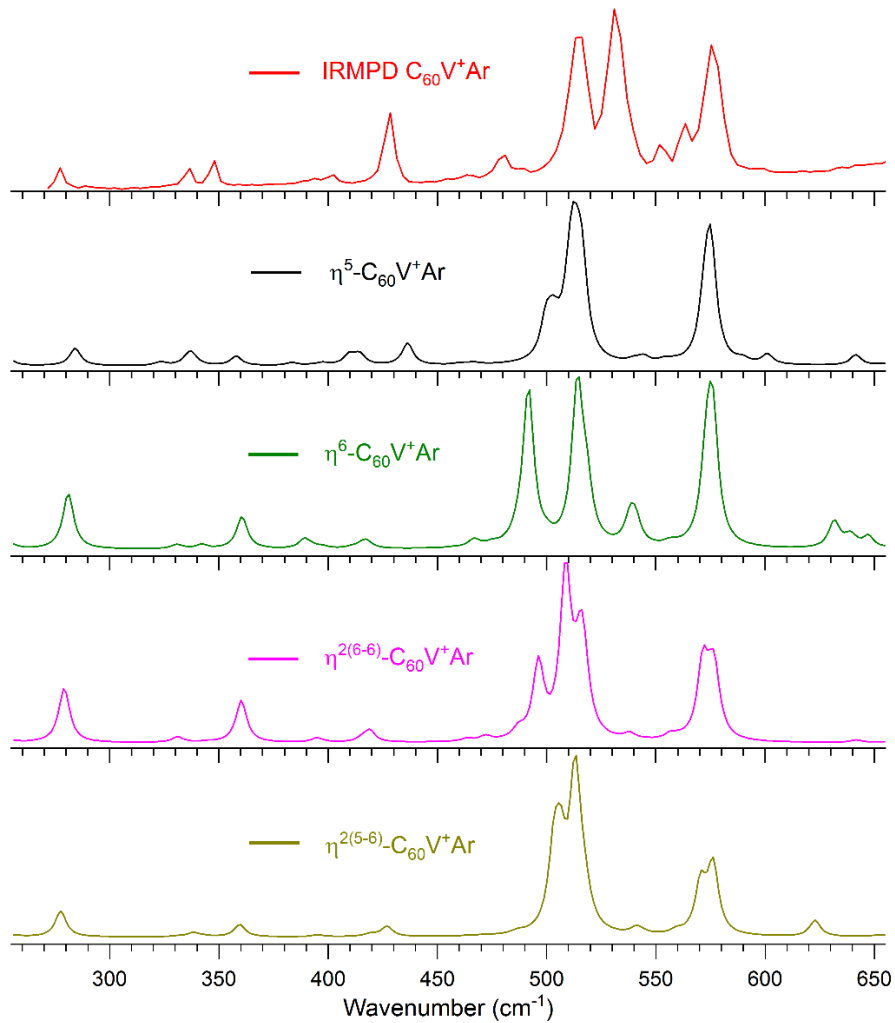

**Figure S5.** Zoom of the IRMPD spectrum of $C_{60}V^+Ar$ and its comparison with simulated ones for different isomers in the 250–650 $cm^{-1}$ spectral range. The calculated cosine similarity scores in the 250–500 $cm^{-1}$ spectral range are 0.711 and 0.558 for $\eta^5$ and $\eta^6$ $C_{60}V^+$, respectively. See the vectors of selected vibrational modes in Figure S4.

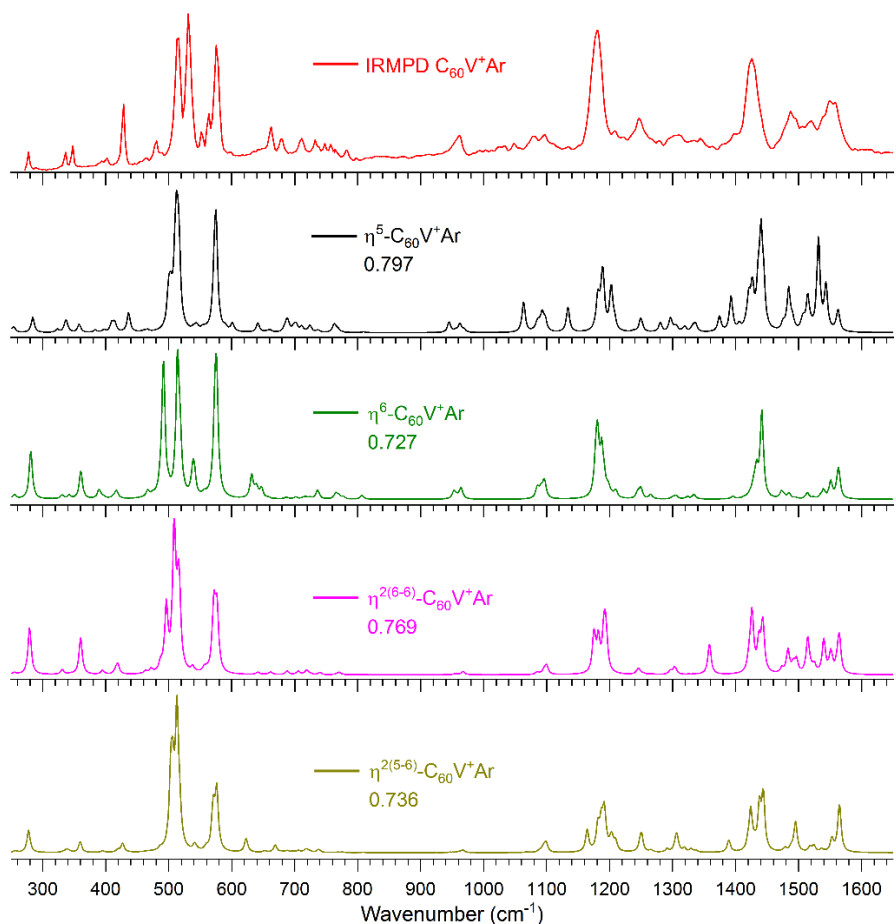

**Figure S6.** Comparison of the experimental spectrum with simulated ones (broadened using Lorentzian line shapes of 6 cm$^{-1}$ full width at half maximum) for the lowest energy spin states (S = 2) in each configuration shown in Figure 2a at the BPW91/6-31G(d) level. The Cosine similarity score of each isomer is indicated for the simulated spectrum.

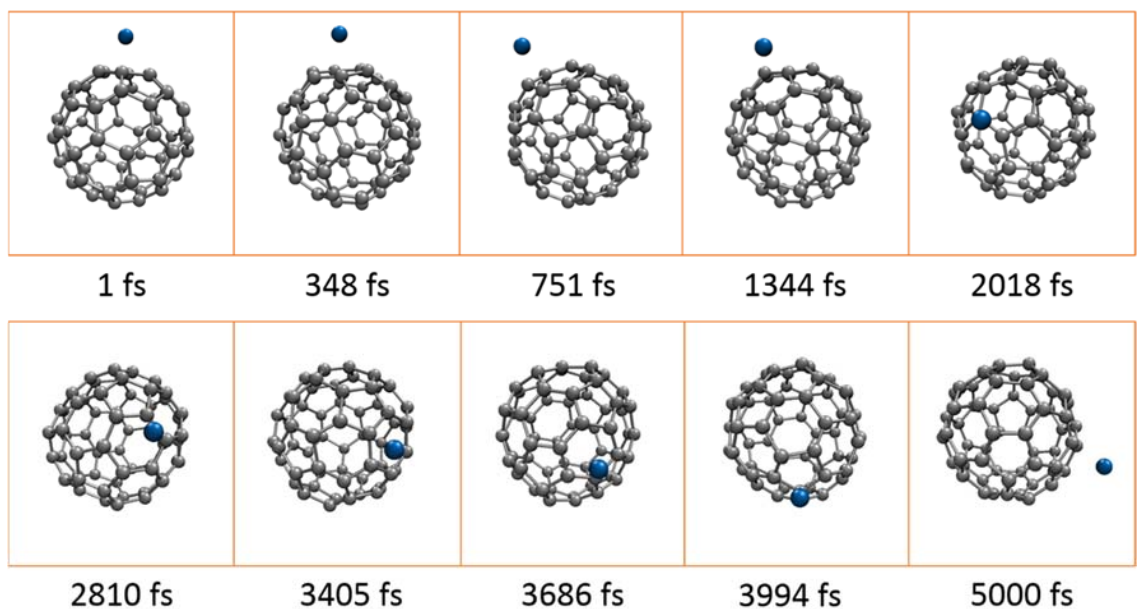

| 1 fs | 348 fs | 751 fs | 1344 fs | 2018 fs |
| 2810 fs | 3405 fs | 3686 fs | 3994 fs | 5000 fs |

**Figure S7.** Snapshots as indicated with grey circles in Figure 3 showing the location changes of V during BOMD simulations at 1000 K.

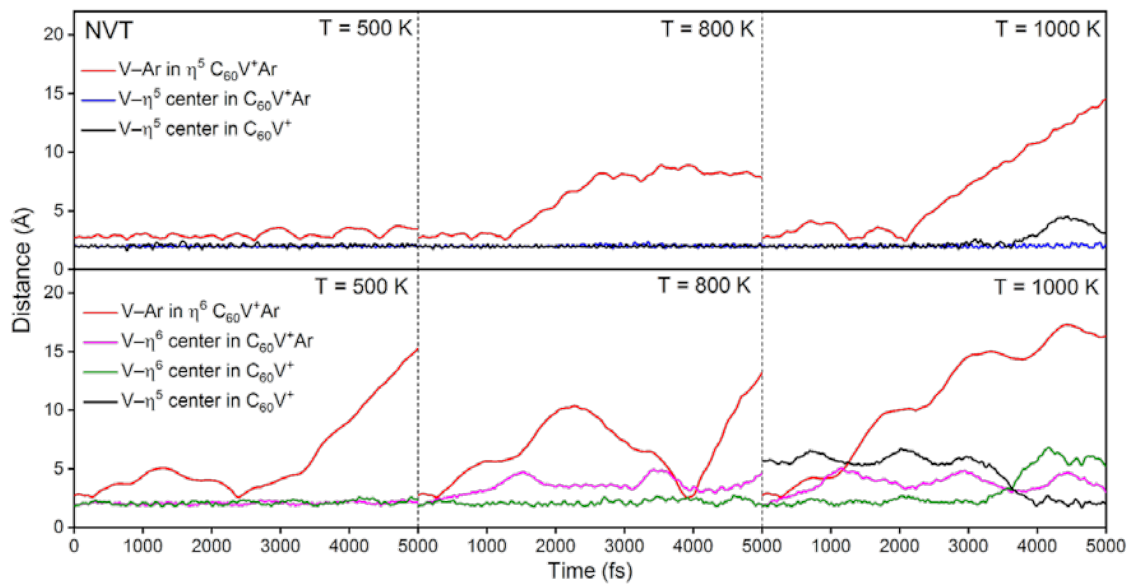

**Figure S8.** Distances between V and Ar (red trace) and between V and the center of the pentagon ($\eta^5$, top) or hexagon ($\eta^6$, bottom) of $C_{60}$ in 5 ps BOMD simulations at 500, 800, and 1000 K with (blue and pink traces) and without (black and green traces) Ar. At low temperature of 500 K, the V–$C_{60}$ distances do not change significantly, and for $\eta^6$ isomer, a significant increase in the V–$C_{60}$ distance is first observed at 800 K. The simulations show that the Ar-tag gets detached from $C_{60}V^+$ in the temperature range of 500–800 K for both $\eta^5$ and $\eta^6$ $C_{60}V^+$, in agreement with their calculated low Ar–$C_{60}V^+$ binding energies of only 3–4 kJ/mol. It is noted that the binding energies and bond lengths of V–$C_{60}$ are 2.63/2.82 eV and 1.917/1.909 Å in the $\eta^5$ $C_{60}V^+$ with and without Ar tagging, respectively. The slightly decreasing binding energy after Ar tagging is consistent with the observation in the BOMD simulations that Ar tagging leads to a larger bond distance of the vanadium to the $C_{60}$ at lower temperatures compared to the non-tagged complex. Similarly, in the case of $\eta^6$ $C_{60}V^+$, the binding energies and bond lengths are 2.56/2.74 eV and 1.884/1.864 Å with and without Ar tagging, respectively.

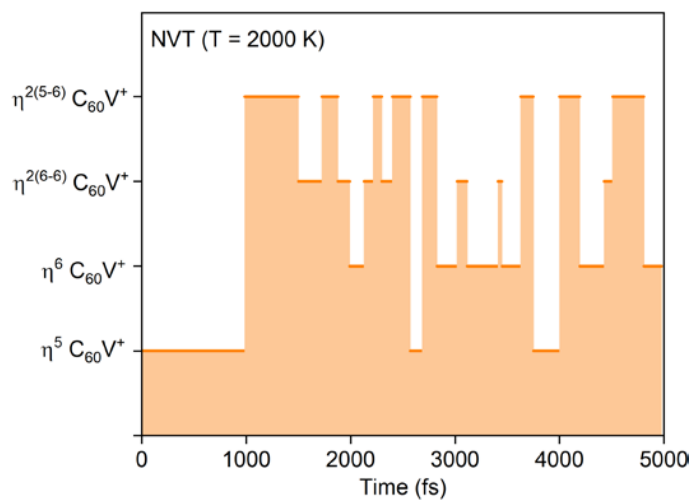

**Figure S9.** Statistics of $C_{60}V^+$ in different isomers along the trajectories in a 5 ps BOMD simulation at 2000 K with $\eta^5$ $C_{60}V^+$ as the starting point. Note that the trajectories after 200 steps are estimated considering the heat balance. The transition time in between different isomers was ignored during the statistical analysis as they only constitute about 7% of the total simulation time.

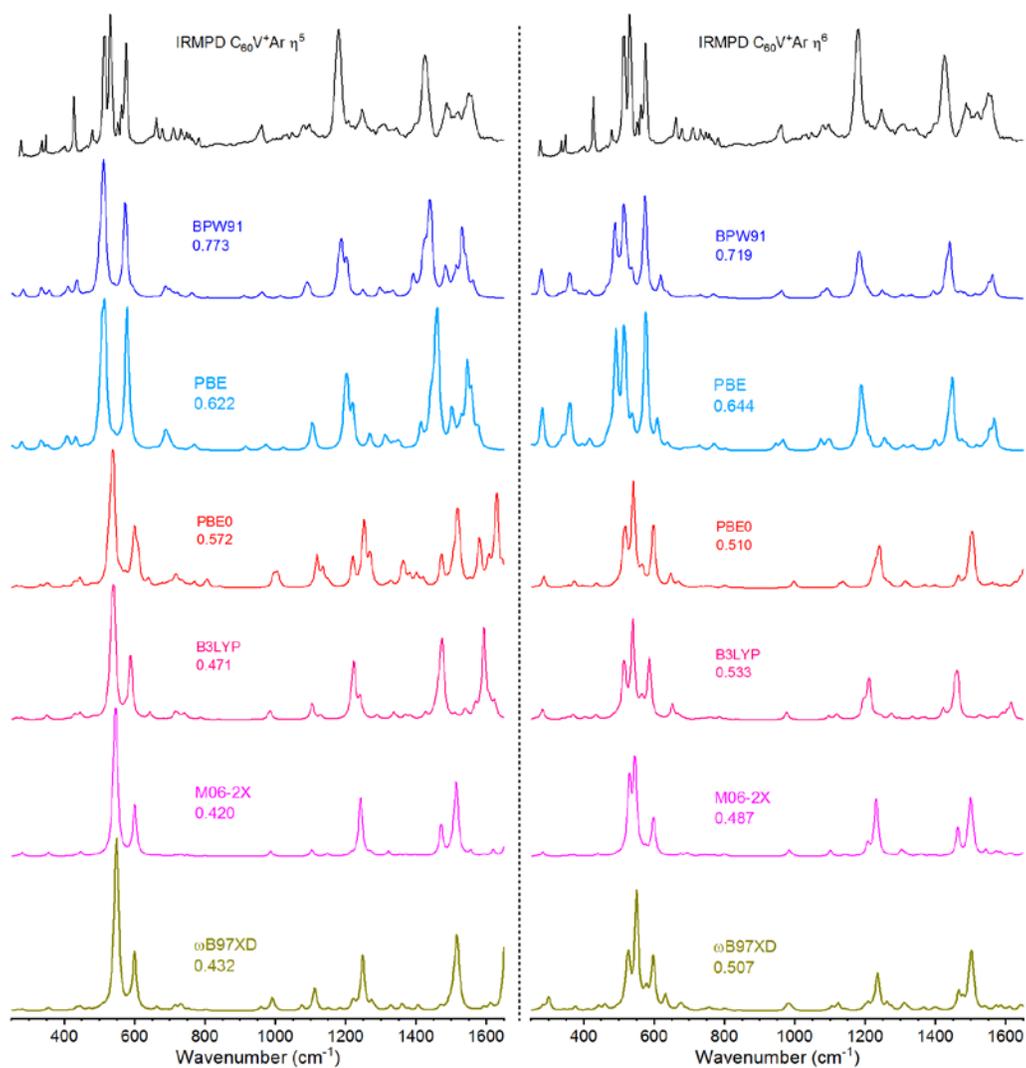

**Figure S10.** Simulated infrared spectra of $\eta^5$ and $\eta^6$ $C_{60}V^+$ (S = 2) without Ar-tag using different functionals and their comparison with experimental IRMPD spectrum. The spectra are convolved using Lorentzian line shapes of 12 $cm^{-1}$ full width at half maximum, and the calculated Cosine similarity scores are indicated. It can be seen that BPW91 gives highest Cosine similarity score and overall best agreement in comparison with experiment without applying any scaling factors.

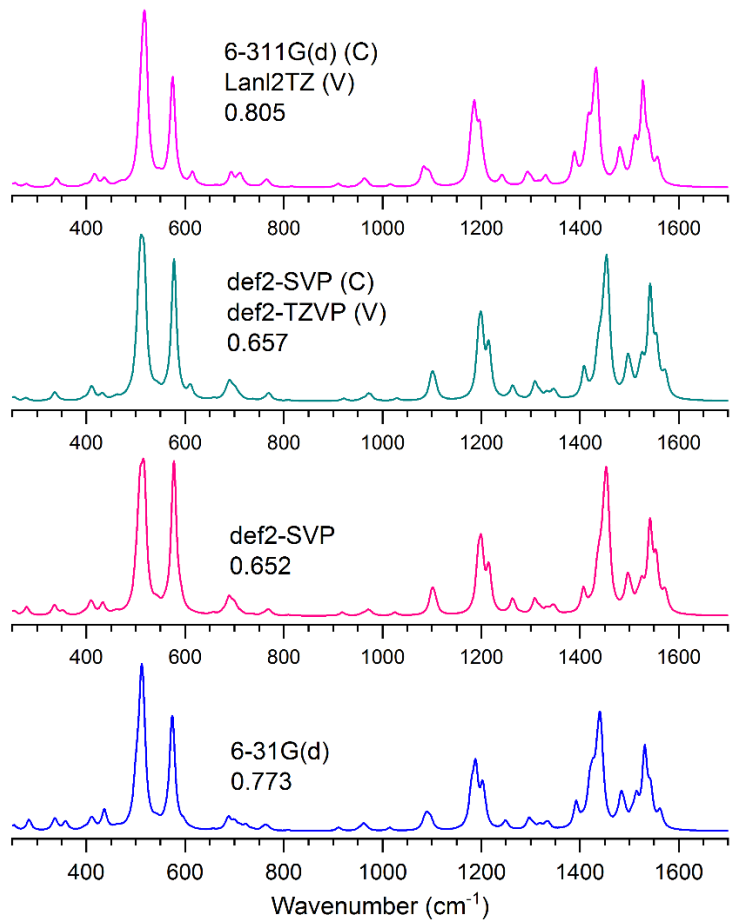

**Figure S11.** Simulated infrared spectra of the $\eta^5$ $C_{60}V^+$ (S = 2) using the BPW91 functional with different basis sets. The spectra are convolved using Lorentzian line shapes of 12 cm$^{-1}$ full width at half maximum, and the calculated Cosine similarity scores are indicated. It shows that the basis sets only have minor effects on the simulated spectra regarding both the band positions and band intensities. Considering both the accuracy and computational cost, 6-31G(d) basis set has been employed.

**Table S1.** The relative stabilities (in kJ/mol) and averaged V–C distances (in Å) of the four isomers of $C_{60}V^+$ with different spin states at various levels of theory. For some isomer and theory level combinations (indicated as "—", the isomers are not stable.

| | | BPW91 | | PBE | | PBE0 | | B3LYP | | M06-2X | | ωB97XD | |
|---|---|---|---|---|---|---|---|---|---|---|---|---|---|
| | | ΔE | V–C | ΔE | V–C | ΔE | V–C | ΔE | V–C | ΔE | V–C | ΔE | V–C |
| $\eta^5$ | S = 0 | 128.0 | 2.140 | 140.2 | 2.153 | 192.5 | 2.207 | 164.5 | 2.243 | 208.2 | 2.289 | 156.2 | 2.234 |
| | S = 1 | 46.9 | 2.250 | 46.4 | 2.267 | 68.2 | 2.291 | 67.0 | 2.328 | 79.0 | 2.510 | 80.7 | 2.320 |
| | **S = 2** | 0 | 2.282 | 0 | 2.306 | 4.9 | 2.384 | 0 | 2.416 | 14.4 | 2.473 | 11.2 | 2.417 |
| $\eta^6$ | S = 0 | 138.0 | 2.220 | 145.7 | 2.229 | 269.0 | 2.314 | 157.8 | 2.350 | 184.2 | 2.338 | 289.8 | 2.151 |
| | S = 1 | 39.3 | 2.246 | 41.3 | 2.263 | 77.7 | 2.370 | 51.4 | 2.585 | 79.1 | 2.346 | 46.1 | 2.466 |
| | **S = 2** | 7.5 | 2.358 | 11.5 | 2.419 | 0 | 2.446 | 0.9 | 2.510 | 0 | 2.494 | 0 | 2.451 |
| $\eta^{2(6\text{-}6)}$ | S = 0 | 207.8 | 1.912 | — | — | 216.2 | 2.036 | 177.2 | 2.056 | 266.0 | 2.024 | 167.6 | 2.064 |
| | S = 1 | — | — | — | — | 77.3 | 2.025 | 55.1 | 2.078 | 86.0 | 2.114 | 46.0 | 2.039 |
| | **S = 2** | 31.4 | 2.181 | 33.6 | 2.202 | 11.4 | 2.254 | 3.59 | 2.290 | 33.0 | 2.304 | 10.2 | 2.274 |
| $\eta^{2(5\text{-}6)}$ | S = 0 | — | — | — | — | — | — | — | — | — | — | — | — |
| | S = 1 | 91.5 | 2.077 | 114.2 | — | — | — | 54.5 | 2.068 | 92.1 | — | — | — |
| | **S = 2** | 35.4 | 2.174 | 40.5 | 2.189 | — | — | — | — | — | — | — | — |

**Table S2.** Cartesian coordinates of the four C$_{60}$V$^+$ (S = 2) isomers with(out) Ar-tag at the BPW91/6-31G(d) level.

| $\eta^5$-C$_{60}$V$^+$ | | | | $\eta^5$-C$_{60}$V$^+$Ar | | | |
|---|---|---|---|---|---|---|---|
| C | 1.51839500 | 2.14801400 | -2.18537400 | C | 1.55186200 | 2.45182100 | -1.40032000 |
| C | 0.27966300 | 1.95887400 | -2.92634500 | C | 0.32919300 | 2.67671700 | -2.16640100 |
| C | -0.91599000 | 2.52091800 | -2.44848100 | C | -0.79956900 | 3.21820400 | -1.53458700 |
| C | -0.91648400 | 3.29708900 | -1.21362000 | C | -0.75311900 | 3.56744400 | -0.11596100 |
| C | 0.27586500 | 3.48608700 | -0.49848300 | C | 0.42195200 | 3.36133000 | 0.62283200 |
| C | 1.51520600 | 2.89217500 | -0.98893500 | C | 1.59842700 | 2.79303300 | -0.02502000 |
| C | 2.28518300 | 0.91246800 | -2.27977700 | C | 2.15466900 | 1.22848600 | -1.87640000 |
| C | 1.52403200 | -0.04132800 | -3.05685100 | C | 1.30151300 | 0.66985100 | -2.92063800 |
| C | 0.28179300 | 0.60522300 | -3.46385300 | C | 0.17665600 | 1.57599000 | -3.10712100 |
| C | -0.91046700 | -0.13345500 | -3.50482800 | C | -1.10018200 | 1.05811100 | -3.38100400 |
| C | -2.15263100 | 1.75642700 | -2.49723100 | C | -2.12499100 | 2.69087700 | -1.82406600 |
| C | -2.15667400 | 3.01230700 | -0.50309900 | C | -2.04852500 | 3.25265700 | 0.46498100 |
| C | -2.15603800 | 2.91937700 | 0.89755900 | C | -2.12306400 | 2.72912800 | 1.76886100 |
| C | -0.91490000 | 3.10672900 | 1.63817700 | C | -0.90522600 | 2.50907700 | 2.53341100 |
| C | 0.27647400 | 3.38942500 | 0.95373100 | C | 0.34489900 | 2.82181900 | 1.97359900 |
| C | 1.51710700 | 2.73656300 | 1.36085500 | C | 1.47105300 | 1.91848400 | 2.15989700 |
| C | 2.28628900 | 2.45523800 | 0.16287700 | C | 2.24442000 | 1.91384600 | 0.92471600 |
| C | 3.02539200 | 1.25201800 | 0.08252600 | C | 2.83538100 | 0.70443100 | 0.47507700 |
| C | 3.03437800 | 0.46807000 | -1.15918700 | C | 2.78528700 | 0.35484400 | -0.95579500 |
| C | 1.52327800 | -1.41433100 | -2.70544600 | C | 1.11075200 | -0.71939100 | -3.01950300 |
| C | 2.28719900 | -1.88540300 | -1.57360700 | C | 1.76376500 | -1.62789500 | -2.08296200 |
| C | 3.03079900 | -0.96094100 | -0.79891800 | C | 2.58460200 | -1.09341300 | -1.05993100 |
| C | 3.02968400 | -1.05897100 | 0.66432300 | C | 2.51202800 | -1.63977300 | 0.30443300 |
| C | 3.03601100 | 0.30962900 | 1.20993300 | C | 2.65681400 | -0.52804400 | 1.25193200 |
| C | 2.28472900 | 0.60296600 | 2.37836700 | C | 1.91326700 | -0.51379900 | 2.45551200 |
| C | 1.51999400 | 1.83910000 | 2.44936000 | C | 1.30530300 | 0.72995000 | 2.89846600 |
| C | 0.28112000 | 1.55322700 | 3.15878600 | C | 0.00969300 | 0.40975600 | 3.48845800 |
| C | -0.91436600 | 2.17382500 | 2.76008200 | C | -1.07401200 | 1.28153700 | 3.30259500 |
| C | -2.92116700 | 2.05980100 | -1.29773400 | C | -2.89555500 | 2.70792900 | -0.58753800 |
| C | -2.91711100 | -0.90560000 | 2.26111800 | C | -3.43476300 | -1.18630100 | 1.86092200 |
| C | -3.65447700 | -0.46211700 | 1.15125500 | C | -4.05822100 | -0.32340300 | 0.94510200 |
| C | -3.65502900 | -1.23878800 | -0.08204500 | C | -4.10391800 | -0.67048800 | -0.46919800 |
| C | -2.92044200 | -2.43286200 | -0.16119600 | C | -3.52326000 | -1.86772000 | -0.91801900 |
| C | -2.15252100 | -2.89088000 | 0.98884400 | C | -2.87272200 | -2.76178700 | 0.03016100 |
| C | -0.90964600 | -1.95330700 | 2.91397000 | C | -1.61140800 | -2.65038800 | 2.15963900 |
| C | -0.91005800 | -0.59693000 | 3.45726500 | C | -1.46015200 | -1.54308000 | 3.09767400 |
| C | -2.14966100 | 0.04647500 | 3.05263700 | C | -2.58872400 | -0.63961600 | 2.91387500 |
| C | -2.15099200 | 1.41048600 | 2.70736600 | C | -2.39896800 | 0.74773100 | 3.01410100 |
| C | -2.92029700 | 1.86988600 | 1.55944200 | C | -3.04783400 | 1.64366200 | 2.06559200 |
| C | -3.65516100 | 0.95024400 | 0.79390600 | C | -3.86082400 | 1.11688300 | 1.04908600 |
| C | -3.65485800 | -0.30579100 | -1.20150800 | C | -3.93588200 | 0.55498400 | -1.23966600 |
| C | -2.91651000 | -0.59857700 | -2.35943600 | C | -3.19646600 | 0.53981600 | -2.43336600 |
| C | -2.15113300 | -1.83569000 | -2.44088500 | C | -2.59174000 | -0.70166800 | -2.89750800 |
| C | -2.15262500 | -2.73569400 | -1.36157500 | C | -2.75253100 | -1.88303800 | -2.15450400 |
| C | -0.91456900 | -3.38587800 | -0.95495400 | C | -1.62244400 | -2.78225500 | -1.96797500 |
| C | -0.91436400 | -3.48180900 | 0.49914200 | C | -1.69868900 | -3.32763800 | -0.61459200 |
| C | 0.28000400 | -3.30587000 | 1.21671400 | C | -0.52517300 | -3.54964600 | 0.12278100 |
| C | 0.28161700 | -2.52637000 | 2.44630900 | C | -0.48029600 | -3.20423000 | 1.53733600 |
| C | 0.28286800 | 0.14042500 | 3.51429000 | C | -0.18695800 | -1.02843600 | 3.38418800 |
| C | 1.52494600 | -0.44610300 | 3.02655800 | C | 0.98779600 | -1.59728900 | 2.73018000 |
| C | 1.52309300 | -1.75894900 | 2.49461000 | C | 0.84335600 | -2.67181400 | 1.82752100 |
| C | 2.28760000 | -2.07740100 | 1.30991900 | C | 1.61175300 | -2.69878200 | 0.59171700 |
| C | 1.51758000 | -3.01806700 | 0.50348900 | C | 0.77211900 | -3.23193700 | -0.46111000 |
| C | 1.51764100 | -2.92611000 | -0.89919700 | C | 0.84492400 | -2.69909800 | -1.77128700 |
| C | 0.27988000 | -3.51809900 | -1.64321200 | C | -0.37813200 | -2.47867300 | -2.53859200 |
| C | 0.28171300 | -2.18073600 | -2.75892400 | C | -0.21202600 | -1.25608600 | -3.31115200 |
| C | -0.90980600 | -3.16849300 | -3.14541300 | C | -1.29745100 | -0.38187300 | -3.48446400 |
| C | -3.65607400 | 1.04720500 | -0.66040200 | C | -3.78452400 | 1.65943800 | -0.30105700 |
| C | -2.15109300 | 0.45011800 | -3.02022000 | C | -2.27222300 | 1.62621900 | -2.72963100 |
| C | -2.15155600 | -1.14249900 | 2.17767300 | C | -2.82907100 | -2.42665600 | 1.39633800 |
| V | 4.94013000 | -0.00089700 | -0.00009100 | V | 4.57645600 | -0.68577800 | -0.00135700 |
| | | | | Ar | 6.95802600 | 0.59966000 | 0.00119500 |

S14

| η⁶-C₆₀V⁺ | | | | η⁶-C₆₀V⁺Ar | | |
|---|---|---|---|---|---|---|
| C | -1.347567000 | -1.899721000 | -2.522905000 | C | -0.65140700 | -2.90151500 | 1.60836100 |
| C | 0.001112000 | -2.442374000 | -2.579571000 | C | 0.68986600 | -2.81887900 | 2.16636700 |
| C | 0.488439000 | -3.223647000 | -1.516850000 | C | 1.01658000 | -1.78224100 | 3.05868000 |
| C | -0.354561000 | -3.493052000 | -0.360451000 | C | 0.01437700 | -0.79217500 | 3.42773700 |
| C | -1.660246000 | -2.976065000 | -0.307527000 | C | -1.28419500 | -0.87505700 | 2.89598600 |
| C | -2.163586000 | -2.174788000 | -1.414620000 | C | -1.61957200 | -1.95551400 | 1.97883900 |
| C | -1.324153000 | -0.562229000 | -3.118583000 | C | -0.53847000 | -3.34738400 | 0.21832100 |
| C | 0.040763000 | -0.288047000 | -3.538776000 | C | 0.87366900 | -3.53780400 | -0.07288400 |
| C | 0.859195000 | -1.447621000 | -3.207668000 | C | 1.63250200 | -3.21193000 | 1.12835700 |
| C | 2.174519000 | -1.270287000 | -2.752179000 | C | 2.86802000 | -2.55543900 | 1.01997100 |
| C | 1.854144000 | -3.040082000 | -1.044643000 | C | 2.30019600 | -1.10235800 | 2.94715600 |
| C | 0.487182000 | -3.465894000 | 0.827711000 | C | 0.67516600 | 5.00091300 | 3.53452000 |
| C | -0.001822000 | -2.919721000 | 2.028092000 | C | 0.01897000 | 1.66726300 | 3.10251400 |
| C | -1.350126000 | -2.375402000 | 2.081819000 | C | -1.32185400 | 1.58215900 | 2.54406500 |
| C | -2.164894000 | -2.418219000 | 0.939669000 | C | -1.96301200 | 0.33665000 | 2.45732400 |
| C | -2.969920000 | -1.250311000 | 0.605498000 | C | -2.70856300 | 0.01091600 | 1.24721900 |
| C | -2.969433000 | -1.100029000 | -0.849576000 | C | -2.49611100 | -1.40623000 | 0.95225600 |
| C | -2.948723000 | 0.206632000 | -1.424455000 | C | -2.38807400 | -1.83349200 | -0.40391900 |
| C | -2.119338000 | 0.474229000 | -2.597130000 | C | -1.39837800 | -2.84213800 | -0.77386800 |
| C | 0.571953000 | 1.007195000 | -3.407118000 | C | 1.38295100 | -3.19763100 | -1.33873500 |
| C | -0.244146000 | 2.075980000 | -2.849509000 | C | 0.49733900 | -2.65617200 | -2.35921200 |
| C | -1.566377000 | 1.814712000 | -2.449066000 | C | -0.86985600 | -2.47934700 | -2.08301700 |
| C | -2.066582000 | 2.392352000 | -1.209883000 | C | -1.54290900 | -1.27191200 | -2.54073400 |
| C | -2.910111000 | 1.397901000 | -0.568917000 | C | -2.47224200 | -0.86411500 | -1.50031500 |
| C | -2.911628000 | 1.252677000 | 0.839641000 | C | -2.67696300 | 0.50672300 | -1.21584600 |
| C | -2.950740000 | -0.088429000 | 1.434628000 | C | -2.80590100 | 0.94923300 | 0.17540400 |
| C | -2.122701000 | -0.065968000 | 2.638101000 | C | -2.16021000 | 2.25501200 | 0.28978100 |
| C | -1.327652000 | -1.187469000 | 2.938000000 | C | -1.41977200 | 2.54966500 | 1.44890800 |
| C | 1.853414000 | -3.189954000 | 0.404705000 | C | 2.08913200 | 0.30904200 | 3.24167900 |
| C | 1.935057000 | 0.565397000 | 3.120010000 | C | 1.78607100 | 3.37791500 | -0.21932700 |
| C | 2.720791000 | -0.462996000 | 2.575213000 | C | 2.64014600 | 2.85012900 | 0.76262400 |
| C | 3.566341000 | -0.201490000 | 1.418898000 | C | 3.64449500 | 1.85996400 | 0.40043200 |
| C | 3.592713000 | 1.080892000 | 0.844663000 | C | 3.75704500 | 1.42953100 | -0.93260500 |
| C | 2.775031000 | 2.146768000 | 1.407792000 | C | 2.86950200 | 1.97510800 | -1.95082500 |
| C | 0.612822000 | 2.440361000 | 2.579829000 | C | 0.56124200 | 2.84828900 | -2.16259100 |
| C | -0.247268000 | 1.449026000 | 3.211587000 | C | -0.38408300 | 3.24506800 | -1.12759800 |
| C | 0.568381000 | 0.289419000 | 3.539937000 | C | 0.37225000 | 3.56763200 | 0.07345100 |
| C | 0.037224000 | -1.005711000 | 3.407010000 | C | -0.13725900 | 3.22648900 | 1.33901300 |
| C | 0.856128000 | -2.073937000 | 2.846756000 | C | 0.75122400 | 2.68296000 | 2.35883200 |
| C | 2.170876000 | -1.805447000 | 2.436516000 | C | 2.11264300 | 2.49674800 | 2.07463900 |
| C | 3.540547000 | -1.381381000 | 0.565261000 | C | 3.73945600 | 0.89511300 | 1.48776900 |
| C | 3.541628000 | -1.237297000 | -0.832169000 | C | 3.94303900 | -0.46569400 | 1.20373100 |
| C | 3.568510000 | 0.092314000 | -1.427532000 | C | 4.05868800 | -0.91172200 | -0.17818500 |
| C | 3.593746000 | 1.230534000 | -0.603705000 | C | 3.96775600 | 0.01916000 | -1.22700800 |
| C | 2.776889000 | 2.389677000 | -0.938036000 | C | 3.21082000 | -0.30891800 | -2.42770000 |
| C | 2.270406000 | 2.956139000 | 0.305820000 | C | 2.53158000 | 0.90038100 | -2.87563600 |
| C | 0.972935000 | 3.487089000 | 0.360217000 | C | 1.24077000 | 0.82132700 | -3.42031600 |
| C | 0.128097000 | 3.224269000 | 1.518161000 | C | 0.23682900 | 1.81338400 | -3.05744900 |
| C | -1.570034000 | 1.276822000 | 2.766916000 | C | -1.62826900 | 2.59868100 | -1.02328100 |
| C | -2.067962000 | 2.095502000 | 1.670359000 | C | -1.96093700 | 1.53129500 | -1.95635700 |
| C | -1.236081000 | 3.042910000 | 1.047493000 | C | -1.04564900 | 1.13562300 | -2.94822400 |
| C | -1.235357000 | 3.193500000 | -0.406399000 | C | -0.83453400 | -0.28003600 | -3.24321300 |
| C | 0.129413000 | 3.466882000 | -0.828185000 | C | 0.57779400 | -0.47154100 | -3.53368000 |
| C | 0.615836000 | 2.917650000 | -2.027934000 | C | 1.23120700 | -1.63805700 | -3.09881200 |
| C | 1.965095000 | 2.371004000 | -2.083672000 | C | 2.57322300 | -1.55514700 | -2.53834100 |
| C | 1.938459000 | 1.190611000 | -2.937674000 | C | 2.66708200 | -2.52006600 | -1.45044700 |
| C | 2.724577000 | 0.072489000 | -2.614132000 | C | 3.39523900 | -2.20297800 | -0.29223600 |
| C | 2.679127000 | -2.373372000 | 1.193949000 | C | 2.79362500 | 1.28831400 | 2.52306600 |
| C | 2.680245000 | -2.080090000 | -1.650297000 | C | 3.20792300 | -1.48136300 | 1.94516100 |
| C | 1.962534000 | 1.895178000 | 2.524960000 | C | 1.90298700 | 2.93204400 | -1.60171100 |
| V | -4.806032000 | 0.001283000 | -0.002621000 | V | -4.45302700 | -0.71643000 | -0.04597600 |
| | | | | Ar | -6.81775000 | 0.61681200 | 0.03938400 |

| $\eta^{2(6\text{-}6)}$-C$_{60}$V$^+$ | | | |
|---|---|---|---|
| C | 0.388311000 | 2.300065000 | 2.622469000 |
| C | -1.068165000 | 2.302458000 | 2.626067000 |
| C | -1.767621000 | 1.160637000 | 3.049668000 |
| C | -1.037822000 | -0.021116000 | 3.492286000 |
| C | 0.365430000 | -0.021009000 | 3.493514000 |
| C | 1.091913000 | 1.161102000 | 3.049620000 |
| C | 0.837866000 | 3.037050000 | 1.449133000 |
| C | -0.338997000 | 3.489172000 | 0.724381000 |
| C | -1.518789000 | 3.038163000 | 1.452090000 |
| C | -2.651535000 | 2.605505000 | 0.745462000 |
| C | -2.946447000 | 0.714285000 | 2.319379000 |
| C | -1.767466000 | -1.197546000 | 3.035491000 |
| C | -1.067775000 | -2.334081000 | 2.598312000 |
| C | 0.388733000 | -2.331406000 | 2.594757000 |
| C | 1.092175000 | -1.197532000 | 3.035449000 |
| C | 2.265224000 | -0.744602000 | 2.304424000 |
| C | 2.265135000 | 0.717185000 | 2.313292000 |
| C | 2.706907000 | 1.437635000 | 1.190095000 |
| C | 1.977550000 | 2.610696000 | 0.742674000 |
| C | -0.339384000 | 3.497624000 | -0.682355000 |
| C | 0.836986000 | 3.054315000 | -1.413232000 |
| C | 1.977090000 | 2.619497000 | -0.712625000 |
| C | 2.706335000 | 1.452006000 | -1.174588000 |
| C | 3.218624000 | 0.733475000 | 0.003357000 |
| C | 3.218485000 | -0.732795000 | -0.005486000 |
| C | 2.707186000 | -1.451431000 | 1.172668000 |
| C | 1.977955000 | -2.619102000 | 0.711149000 |
| C | 0.838332000 | -3.054060000 | 1.412496000 |
| C | -2.946304000 | -0.742667000 | 2.310567000 |
| C | -1.519108000 | -3.038356000 | -1.451352000 |
| C | -2.651517000 | -2.605884000 | -0.744027000 |
| C | -3.381367000 | -1.424302000 | -1.185700000 |
| C | -2.947692000 | -0.714737000 | -2.317791000 |
| C | -1.769257000 | -1.160894000 | -3.048813000 |
| C | 0.387117000 | -2.299926000 | -2.622957000 |
| C | 0.837570000 | -3.036906000 | -1.449874000 |
| C | -0.338789000 | -3.489153000 | -0.724383000 |
| C | -0.338392000 | -3.497661000 | 0.682352000 |
| C | -1.518185000 | -3.055602000 | 1.415434000 |
| C | -2.651101000 | -2.614697000 | 0.714001000 |
| C | -3.831290000 | -0.703052000 | -0.003168000 |
| C | -3.831391000 | 0.702444000 | 0.005313000 |
| C | -3.381605000 | 1.438042000 | -1.168424000 |
| C | -2.947794000 | 0.742111000 | -2.308966000 |
| C | -1.769513000 | 1.197304000 | -3.034644000 |
| C | -1.039984000 | 0.021000000 | -3.491869000 |
| C | 0.363303000 | 0.021139000 | -3.493935000 |
| C | 1.090217000 | -1.160843000 | -3.050506000 |
| C | 1.977596000 | -2.610342000 | -0.744158000 |
| C | 2.706545000 | -1.437150000 | -1.192040000 |
| C | 2.263839000 | -0.716656000 | -2.314793000 |
| C | 2.263721000 | 0.745018000 | -2.305936000 |
| C | 1.090080000 | 1.197787000 | -3.036328000 |
| C | 0.386748000 | 2.331531000 | -2.595250000 |
| C | -1.069775000 | 2.333958000 | -2.597847000 |
| C | -1.519550000 | 3.055368000 | -1.414710000 |
| C | -2.651897000 | 2.614276000 | -0.712572000 |
| C | -3.380641000 | -1.438558000 | 1.170252000 |
| C | -3.380896000 | 1.423762000 | 1.187525000 |
| C | -1.069368000 | -2.302591000 | -2.625576000 |
| V | 5.272790000 | -0.000012000 | 0.000060000 |

| $\eta^{2(6\text{-}6)}$-C$_{60}$V$^+$Ar | | | |
|---|---|---|---|
| C | -0.02934100 | -2.31608200 | 2.60882000 |
| C | 1.42739600 | -2.31841500 | 2.61170900 |
| C | 2.12693700 | -1.17923200 | 3.04240000 |
| C | 1.39743600 | -0.00000700 | 3.49195800 |
| C | -0.00587800 | -0.00000700 | 3.49353600 |
| C | -0.73253700 | -1.17951100 | 3.04309700 |
| C | -0.47913500 | -3.04528100 | 1.43093400 |
| C | 0.69755000 | -3.49327500 | 0.70321200 |
| C | 1.87768200 | -3.04672200 | 1.43316600 |
| C | 3.01059200 | -2.60984200 | 0.72915300 |
| C | 3.30561600 | -0.72852400 | 2.31471300 |
| C | 2.12693700 | 1.17921900 | 3.04240500 |
| C | 1.42739600 | 2.31840400 | 2.61171800 |
| C | -0.02934100 | 2.31607100 | 2.60882900 |
| C | -0.73253700 | 1.17949800 | 3.04310200 |
| C | -1.90651400 | 0.73084600 | 2.31048200 |
| C | -1.90651400 | -0.73085500 | 2.31047900 |
| C | -2.34840000 | -1.44356900 | 1.18231100 |
| C | -1.61890400 | -2.61378400 | 0.72756100 |
| C | 0.69758400 | -3.49327100 | -0.70322400 |
| C | -0.47906600 | -3.04527400 | -1.43100200 |
| C | -1.61886800 | -2.61378100 | -0.72768300 |
| C | -2.34834300 | -1.44356400 | -1.18246400 |
| C | -2.86048900 | -0.73229100 | -0.00008800 |
| C | -2.86048900 | 0.73229200 | -0.00008500 |
| C | -2.34840000 | 1.44356400 | 1.18231700 |
| C | -1.61890400 | 2.61378100 | 0.72757200 |
| C | -0.47913600 | 3.04527500 | 1.43094700 |
| C | 3.30561600 | 0.72851400 | 2.31471600 |
| C | 1.87775300 | 3.04672200 | -1.43310600 |
| C | 3.01062800 | 2.60984200 | -0.72903800 |
| C | 3.73996100 | 1.43093800 | -1.17814000 |
| C | 3.30572900 | 0.72852400 | -2.31458400 |
| C | 2.12708600 | 1.17923200 | -3.04232800 |
| C | -0.02921300 | 2.31608100 | -2.60885200 |
| C | -0.47906600 | 3.04528000 | -1.43098900 |
| C | 0.69758400 | 3.49327400 | -0.70320900 |
| C | 0.69755000 | 3.49327200 | 0.70322700 |
| C | 1.87768300 | 3.04671600 | 1.43317900 |
| C | 3.01059200 | 2.60983800 | 0.72916400 |
| C | 4.18999300 | 0.70265500 | 0.00008800 |
| C | 4.18999300 | -0.70265500 | 0.00008500 |
| C | 3.73996100 | -1.43093400 | -1.17814600 |
| C | 3.30572900 | -0.72851500 | -2.31458700 |
| C | 2.12708600 | -1.17921900 | -3.04233300 |
| C | 1.39760600 | 0.00000700 | -3.49192200 |
| C | -0.00570800 | 0.00000000 | -3.49356700 |
| C | -0.73238900 | 1.17951100 | -3.04316400 |
| C | -1.61886800 | 2.61378400 | -0.72767200 |
| C | -2.34834300 | 1.44356900 | -1.18245800 |
| C | -1.90640100 | 0.73085600 | -2.31060300 |
| C | -1.90640100 | -0.73084000 | -2.31060600 |
| C | -0.73238900 | -1.17949800 | -3.04316900 |
| C | -0.02921300 | -2.31607000 | -2.60886200 |
| C | 1.42752400 | -2.31840400 | -2.61168000 |
| C | 1.87775300 | -3.04671600 | -1.43311900 |
| C | 3.01062700 | -2.60983900 | -0.72904900 |
| C | 3.73990300 | 1.43093300 | 1.17829700 |
| C | 3.73990300 | -1.43093900 | 1.17829100 |
| C | 1.42752400 | 2.31841500 | -2.61167000 |
| V | -4.92503900 | 0.00000000 | -0.00005800 |
| Ar | -7.61626800 | 0.00000000 | 0.00005200 |

| η²(5-6)-C₆₀V⁺ | | | | η²(5-6)-C₆₀V⁺Ar | | |
|---|---|---|---|---|---|---|
| C | 0.066919000 | 3.515047000 | 0.401143000 | C | -0.39075300 | -3.49074300 | 0.61893300 |
| C | -1.297352000 | 3.321209000 | 0.869578000 | C | -1.76226700 | -3.39511200 | 0.14027400 |
| C | -1.533178000 | 2.639771000 | 2.076082000 | C | -2.01375400 | -3.08050400 | -1.20621600 |
| C | -0.412343000 | 2.133169000 | 2.855695000 | C | -0.90195700 | -2.85397100 | -2.11903200 |
| C | 0.902755000 | 2.329102000 | 2.406944000 | C | 0.41973700 | -2.95596100 | -1.65790800 |
| C | 1.147241000 | 3.032847000 | 1.156475000 | C | 0.68044800 | -3.27998100 | -0.26291200 |
| C | 0.067395000 | 3.376882000 | -1.053074000 | C | -0.34696100 | -2.94221800 | 1.97167500 |
| C | -1.296712000 | 3.098708000 | -1.477815000 | C | -1.69115700 | -2.50948300 | 2.32411300 |
| C | -2.140645000 | 3.063263000 | -0.290535000 | C | -2.56629000 | -2.78816200 | 1.19310300 |
| C | -3.185830000 | 2.130126000 | -0.202258000 | C | -3.58977400 | -1.88633300 | 0.86125300 |
| C | -2.620302000 | 1.674673000 | 2.167953000 | C | -3.07773000 | -2.14666800 | -1.55005800 |
| C | -0.806112000 | 0.850214000 | 3.426876000 | C | -1.27808000 | -1.77630000 | -3.02660800 |
| C | 0.128549000 | -0.193872000 | 3.527607000 | C | -0.32012100 | -0.83600800 | -3.44140400 |
| C | 1.489461000 | 0.005410000 | 3.050115000 | C | 1.04795600 | -0.93420300 | -2.95251000 |
| C | 1.871207000 | 1.244479000 | 2.506816000 | C | 1.41212100 | -1.97720500 | -2.08287300 |
| C | 2.710708000 | 1.285258000 | 1.322351000 | C | 2.28380800 | -1.70531200 | -0.95308000 |
| C | 2.265608000 | 2.381118000 | 0.484824000 | C | 1.83428300 | -2.50097700 | 0.17215300 |
| C | 2.266646000 | 2.248482000 | -0.921576000 | C | 1.87610800 | -1.97071400 | 1.48017200 |
| C | 1.148085000 | 2.762292000 | -1.703854000 | C | 0.76650100 | -2.20211100 | 2.39801300 |
| C | -1.532060000 | 2.203014000 | -2.534885000 | C | -1.87443900 | -1.34218800 | 3.08476800 |
| C | -0.410816000 | 1.558643000 | -3.204091000 | C | -0.71924100 | -0.56997500 | 3.52050300 |
| C | 0.904211000 | 1.836293000 | -2.800516000 | C | 0.57682000 | -0.99379700 | 3.18772100 |
| C | 1.871599000 | 0.751854000 | -2.694157000 | C | 1.56908100 | -0.01740600 | 2.75666700 |
| C | 2.710614000 | 1.014241000 | -1.538116000 | C | 2.36997400 | -0.62739900 | 1.70891000 |
| C | 3.213177000 | -0.065814000 | -0.750300000 | C | 2.87387600 | 0.16584200 | 0.63352000 |
| C | 3.213985000 | 0.076557000 | 0.751242000 | C | 2.83041500 | -0.39992100 | -0.76241800 |
| C | 2.763384000 | -1.212837000 | 1.302525000 | C | 2.40009200 | 0.69155100 | -1.65263500 |
| C | 1.943256000 | -1.241774000 | 2.442076000 | C | 1.54924000 | 0.41956100 | -2.73657600 |
| C | -2.170653000 | 0.568548000 | 3.003564000 | C | -2.62281200 | -1.34110400 | -2.67617300 |
| C | -0.744346000 | -3.386737000 | 1.051990000 | C | -1.04294100 | 2.95815100 | -1.96930800 |
| C | -1.822290000 | -2.759960000 | 1.696998000 | C | -2.15389300 | 2.20819600 | -2.38705100 |
| C | -2.943246000 | -2.250757000 | 0.918196000 | C | -3.26543500 | 1.97986600 | -1.47349500 |
| C | -2.942070000 | -2.382591000 | -0.480869000 | C | -3.22308900 | 2.50690800 | -0.17167500 |
| C | -1.821346000 | -3.029073000 | -1.150025000 | C | -2.06782800 | 3.28095000 | 0.26259000 |
| C | 0.621110000 | -3.322134000 | -0.867602000 | C | 0.37277000 | 3.40194700 | -0.13793400 |
| C | 1.462344000 | -3.062360000 | 0.290443000 | C | 1.17478400 | 2.79416500 | -1.18867700 |
| C | 0.620451000 | -3.100008000 | 1.476055000 | C | 0.30161900 | 2.51793400 | -2.31900600 |
| C | 0.856262000 | -2.203627000 | 2.532597000 | C | 0.48528000 | 1.35006900 | -3.07891600 |
| C | -0.264187000 | -1.557961000 | 3.205094000 | C | -0.66909900 | 0.57553900 | -3.51720600 |
| C | -1.577775000 | -1.830301000 | 2.793239000 | C | -1.96363900 | 0.99657200 | -3.17521800 |
| C | -3.391890000 | -1.007896000 | 1.531322000 | C | -3.76264900 | 0.62900300 | -1.69565000 |
| C | -3.823613000 | 0.058231000 | 0.724738000 | C | -4.19957300 | -0.14699600 | -0.60909500 |
| C | -3.824582000 | -0.078920000 | -0.724760000 | C | -4.15656500 | 0.39933000 | 0.74004000 |
| C | -3.391190000 | -1.277404000 | -1.316456000 | C | -3.67692200 | 1.70231800 | 0.95452100 |
| C | -2.547306000 | -1.239578000 | -2.503324000 | C | -2.80157800 | 1.97817000 | 2.08585800 |
| C | -1.576204000 | -2.321718000 | -2.400796000 | C | -1.80669000 | 2.95364400 | 1.65890300 |
| C | -0.262675000 | -2.132863000 | -2.857129000 | C | -0.48609500 | 2.86037300 | 2.12508200 |
| C | 0.857295000 | -2.640006000 | -2.073825000 | C | 0.62462000 | 3.08664300 | 1.20868300 |
| C | 2.515232000 | -2.130392000 | 0.202306000 | C | 2.20621700 | 1.89388700 | -0.85767100 |
| C | 2.762574000 | -1.435233000 | -1.050035000 | C | 2.47020200 | 1.57850100 | 0.53706700 |
| C | 1.943872000 | -1.678136000 | -2.164938000 | C | 1.68818100 | 2.15573800 | 1.55097200 |
| C | 1.491613000 | -0.567503000 | -2.997516000 | C | 1.23094300 | 1.34494100 | 2.67556300 |
| C | 0.130893000 | -0.853012000 | -3.430118000 | C | -0.10988900 | 1.78606700 | 3.03333700 |
| C | -0.803733000 | 0.191759000 | -3.524300000 | C | -1.06818900 | 0.84352700 | 3.44240000 |
| C | -2.168473000 | -0.005462000 | -3.056615000 | C | -2.43973400 | 0.94269700 | 2.96319000 |
| C | -2.619521000 | 1.237850000 | -2.444498000 | C | -2.93875300 | -0.40854600 | 2.74196600 |
| C | -3.429904000 | 1.201137000 | -1.298681000 | C | -3.77956400 | -0.67508300 | 1.64960600 |
| C | -2.547888000 | -0.747783000 | 2.689826000 | C | -2.95838600 | 0.02106200 | -2.74735700 |
| C | -3.429666000 | 1.423156000 | 1.048615000 | C | -3.84990500 | -1.55960400 | -0.53504100 |
| C | -0.743830000 | -3.524337000 | -0.398117000 | C | -0.99905100 | 3.50496100 | -0.61992800 |
| V | 5.252685000 | 0.006962000 | -0.000653000 | V | 4.89388100 | -0.16007000 | -0.09999400 |
|  |  |  |  | Ar | 7.58995000 | 0.13315700 | 0.08846100 |